\documentclass[11pt]{article}
\usepackage[T1]{fontenc}
\usepackage{textcomp}
\usepackage[sc,noBBpl]{mathpazo}
\usepackage{amsmath,amsfonts,amssymb,amsthm}
\usepackage[mathscr]{eucal}
\usepackage[square,numbers]{natbib}
\newcommand\cM{{\mathcal M}}

\newcommand\cT{{\mathcal T}}

\newcommand\mvector{\boldsymbol}

\newcommand\vd{\mvector{d}}

\newcommand\vp{\mvector{p}}
\newcommand\vq{\mvector{q}}

\newcommand\vv{\mvector{v}}

\newcommand\vx{\mvector{x}}
\newcommand\vy{\mvector{y}}
\newcommand\vz{\mvector{z}}

\newcommand\veta{\mvector{\eta}}
\newcommand\vxi{\mvector{\xi}}

\newcommand\vvarphi{\mvector{\varphi}}

\newcommand\vXi{\mvector{\Xi}}

\newcommand\field{\mathbb}

\newcommand\C{\field{C}}
\newcommand\Z{\field{Z}}
\newcommand\M{\field{M}}
\newcommand\N{\field{N}}
\newcommand\Q{\field{Q}}

\renewcommand\Re{\operatorname{Re}}

\newcommand\rmd{\mathrm{d}\mspace{1mu}}

\newcommand\rmi{\mathrm{i}\mspace{1mu}}

\newcommand\Dt{\frac{\mathrm{d}\phantom{t} }{\mathrm{d}\mspace{1mu} t}}

\newcommand\pder[2]{\dfrac{\partial #1 }{\partial #2}} 
\newcommand\abs[1]{\lvert #1 \rvert}
\newcommand\norm[1]{\lVert #1 \rVert}

\newcommand\mtext[1]{\quad\text{#1}\quad}

\newcommand\defset[2]{\left\{{#1}\;\vert \;\; {#2} \,\right\}}
\usepackage[square]{natbib}
\theoremstyle{plain}
\newtheorem{theorem}{Theorem}
\newtheorem{lemma}{Lemma}
\newtheorem{proposition}{Proposition}
\newtheorem{corollary}{Corollary}
\theoremstyle{definition}

\newtheoremstyle{note}{\topsep}{\topsep}{\slshape}{}{\scshape}{}{ }{}
\theoremstyle{note}

\newtheorem{remark}{Remark}
\numberwithin{equation}{section}
\numberwithin{theorem}{section}
\numberwithin{lemma}{section}
\numberwithin{proposition}{section}
\numberwithin{corollary}{section}
\numberwithin{remark}{section}

\begin{document}
\thispagestyle{empty}
{\flushright{\textsf{submitted to Nonlinearity}}}
\vspace{1em}
\begin{center}
\LARGE\textbf{Necessary conditions for partial and super-integrability of
Hamiltonian systems with homogeneous potential}
\end{center}
\vspace*{0.5em}
\begin{center}
\large Andrzej J.~Maciejewski$^1$, Maria Przybylska$^{2,3}$ and Haruo
Yoshida$^4$
\end{center}
\vspace{2em}
\hspace*{2em}\begin{minipage}{0.8\textwidth}
\small
$^1$Institute of Astronomy,
  University of Zielona G\'ora,
  Podg\'orna 50, \\\quad PL-65--246 Zielona G\'ora, Poland
  (e-mail: maciejka@astro.ia.uz.zgora.pl)\\[0.5em]
$^2$Institut Fourier, UMR 5582 du CNRS,
Universit\'e de Grenoble I, \\\quad
100 rue des Maths,
BP 74, 38402 Saint-Martin d'H\`eres Cedex, France\\[0.5em]
$^3$Toru\'n Centre for Astronomy,
  N.~Copernicus University,
  Gagarina 11,\\\quad PL-87--100 Toru\'n, Poland,
  (e-mail: Maria.Przybylska@astri.uni.torun.pl)\\[0.5em]
$^4$National Astronomical Observatory, 2-21-1 Osawa,  Mitaka, 181-8588 Tokyo,
Japan, (e-mail:
h.yoshida@nao.ac.jp)
\end{minipage}\\[2.5em]
%\maketitle
%\vspace*{1.5em}
{\small \textbf{Abstract.}  
We consider a natural Hamiltonian system of
$n$ degrees of freedom with a homogeneous potential. Such system is
called partially integrable if it admits $1<l<n$ independent and
commuting first integrals, and it is called super-integrable if it
admits $n+l$, $0<l<n$ independent first integrals such that $n$ of them
commute.  We formulate two theorems which give easily computable and
effective necessary conditions for partial and
super-integrability. These conditions are derived in the frame of the
Morales-Ramis theory, i.e., from an analysis of the differential
Galois group of variational equations along a particular solution of
the system.  To illustrate an application of the formulated theorems,
we investigete three and four body problems on a line and the motion in a
radial potential.  } \\[1em] 

\noindent MSC 37J30, 70H06, 53D20 

\section{Introduction}
The fundamental problem in Hamiltonian mechanics is to decide whether
a given system is integrable. Integrability in this context usually means
 the integrability in the Liouville sense~\cite{Arnold:78::},
but it is also important to consider the non-commutative integrability
as it was defined in~\cite{Mishchenko:78::}. Moreover, there exist
examples of systems which are super-integrable, i.e., systems with $n$
degrees of freedom admitting $m>n$ independent first integrals such
that $n$ of them commute. Such super-integrable systems attract much
attention, see e.g.
\cite{MR2151125,MR2105429,Kalnins:01::,Evans:90::}.  On the other
hand, even if a considered system is not integrable, it is anyway
important  to know if it admits one or more first integrals. These
additional first integrals can be used, e.g. to reduce the dimension
of the system.

In this paper, we consider Hamiltonian systems with $n$ degrees of
freedom given by a natural Hamiltonian function
\begin{equation}
 \label{eq:ham}
H = \frac{1}{2}\sum_{i=1}^np_i^2 +V(\vq), 
\end{equation}
where $\vq=(q_1,\ldots,q_n)$ and $\vp=(p_1,\ldots,p_n)$ are canonical
coordinates, and $V$ is a homogeneous function of degree $k\in\Z$. Our
aim is to find necessary conditions for:
\begin{enumerate}
 \item the existence of a meromorphic first integral functionally
independent with $H$; later we call such integral an additional first integral;
\item the existence of commuting functionally independent meromorphic
first integrals $F_1=H$, $F_2,\ldots,F_m$, for $2<m\leq n$;
\item the existence of   functionally independent meromorphic first integrals 
$F_1=H$, $F_2,\ldots,F_{n+m}$, for $0<m<n$, such that $F_1,\ldots,F_n$
commute.
\end{enumerate}
In other words, our goal is  to find necessary conditions for the partial
commutative integrability  and super-integrability.

To formulate the main results of this paper we have to recall a theorem of
J.J.~Morales-Ruiz and J.-P.~Ramis which gives the strongest known
necessary conditions for the integrability in the Liouville sense of
Hamiltonian systems given by a Hamiltonian function of the
form~\eqref{eq:ham}.

The basic assumption of the above mentioned theorem is that Hamilton's
equations generated by~\eqref{eq:ham}, i.e.,
\begin{equation}
 \label{eq:heqs}
 \Dt \vq = \vp, \qquad \Dt \vp = -\pder{V(\vq)}{\vq},
\end{equation}
admit a 
straight line solution  of the form
\begin{equation}
\label{eq:sls}
 \vq(t)= \varphi(t)\vd, \qquad \vp(t)= \dot\varphi(t)\vd, 
\end{equation}
where $\vd$ is a non-zero vector in $\C^n$, and $\varphi(t)$ is a scalar
function.
Such solution exists iff  at  $\vd$ gradient $V'(\vd)$ is
parallel to $\vd$, i.e., $V'(\vd)=\gamma \vd$  for a non-zero $\gamma$. Such
$\vd$
is called the Darboux point of potential $V$. The length of $\vd$ can be fixed
arbitrarily and traditionally $\vd$ is  normalised in such a way that
it satisfies the following non-linear equation 
\begin{equation}
\label{eq:dar}
  V'(\vd) = \vd. 
\end{equation} 
\begin{remark}
 If $V'(\vd)=\gamma \vd$, then $\widetilde \vd:=\alpha \vd$ satisfies
$V'(\widetilde \vd)= \gamma \alpha^{k-2} \widetilde \vd$. Thus, for $k\neq 2$
we can find such $\alpha$ that $V'(\widetilde \vd)=\widetilde \vd$. For $k=2$
we cannot generally assume that for a straight line solution~\eqref{eq:sls}
$\vd$
satisfies~\eqref{eq:dar}. 
\end{remark}
Accepting the above  convention, it is easy to see that~\eqref{eq:sls} is a
solution of~\eqref{eq:heqs} iff $\varphi(t)$ satisfies  $\ddot \varphi = -
\varphi^{k-1}$. For
further
considerations we choose a  phase curve $\Gamma$ corresponding to a non-zero
energy level
\begin{equation}
 \label{eq:eg}
e= \frac{1}{2}\dot\varphi^2 + \frac{1}{k}\varphi^k,   \qquad e\neq 0.
\end{equation} 
The necessary conditions of the Morales-Ramis theorem were obtained by an
analysis of the variational equations along  the considered phase curve. These
equations are of the form
\begin{equation}
 \label{eq:vart}
\ddot \vx = -\varphi(t)^{k-2} V''(\vd)\vx,
\end{equation} 
where  $V''(\vd)$ is the Hessian of $V$ calculated at $\vd$. Let us
assume that  $V''(\vd)$ is diagonalisable. Then, in an appropriate base,
equations~\eqref{eq:vart} split into a direct product of second order equations
\begin{equation}
 \label{eq:varts}
\ddot y_i = -\lambda_i \varphi(t)^{k-2} y_i, \qquad 1\leq i\leq n,
\end{equation}
where $\lambda_1,\ldots, \lambda_n$ are eigenvalues of $V''(\vd)$. One of these
eigenvalues, let us say $\lambda_n$ is  $k-1$. We call this eigenvalue
trivial.

In~\cite{Morales:01::a} J.~J.~Morales-Ruiz
and J.~P.~Ramis proved the following theorem.
\begin{theorem}[Morales-Ramis]
\label{thm:mo2}
If the Hamiltonian system defined by  Hamiltonian \eqref{eq:ham} with a
homogeneous
potential of degree $k\in\Z^\star$ is
integrable in the Liouville sense, then each pair $(k,\lambda_i)$ belongs to an
item of the following list
\begin{equation}
 \label{eq:tab}
\begin{array}{crcr}
\text{case} &  k & \lambda  &\\[0.5em]
\hline
\vbox to 1.3em{}1. &  \pm 2 &  \lambda  & \\[0.7em]
2.  & k & p + \dfrac{k}{2}p(p-1) & \\[0.9em]
3. & k & \dfrac 1 {2}\left(\dfrac {k-1} {k}+p(p+1)k\right)  & \\[0.9em]
4. & 3 &  -\dfrac 1 {24}+\dfrac 1 {6}\left( 1 +3p\right)^2, & -\dfrac 1
{24}+\dfrac 3 {32}\left(  1  +4p\right)^2 \\[0.7em]
 & & -\dfrac 1 {24}+\dfrac 3 {50}\left(  1  +5p\right)^2,  &
-\dfrac1{24}+\dfrac{3}{50}\left(2 +5p\right)^2 \\[0.9em]
5. & 4 & -\dfrac 1 8 +\dfrac{2}{9} \left( 1+ 3p\right)^2 & \\[0.9em]
6. & 5 & -\dfrac 9 {40}+\dfrac 5 {18}\left(1+ 3p\right)^2, & -\dfrac 9
{40}+\dfrac 1 {10}\left(2
  +5p\right)^2\\[0.9em]
7. & -3 &\dfrac {25} {24}-\dfrac 1 {6}\left( 1 +3p\right)^2, & \dfrac {25}
{24}-\dfrac 3 {32}\left(1 +4p\right)^2 \\[0.9em]
 & & \dfrac {25} {24}-\dfrac 3 {50}\left(1+ 5p\right)^2, & \dfrac {25}
{24}-\dfrac{3}{50}\left(2+ 5p\right)^2\\[0.9em]
8. & -4 & \dfrac 9 8-\dfrac{2}{9}\left( 1+ 3p\right)^2 & \\[0.9em]
9. & -5 & \dfrac {49} {40}-\dfrac {5} {18}\left(1+3p\right)^2 ,& \dfrac {49}
{40}-\dfrac {1} {10}(2 +5p)^2\\[0.9em]
\hline
\end{array}
\end{equation}
where $p$ is an integer and $\lambda$ is an arbitrary complex number. 
\end{theorem}
We formulate our main results in two theorems. The first gives necessary
conditions for the partial integrability.
\begin{theorem}
\label{thm:we1}
If a Hamiltonian system defined by Hamiltonian \eqref{eq:ham} with a homogeneous
potential of degree $k\in\Z^\star$ admits $1\leq l \leq n$ functionally
independent and commuting meromorphic first integrals $F_1=H, F_2, \ldots, F_l$,
then at least $l$ pairs of 
$(k,\lambda_i)$ belong to the list~\eqref{eq:tab} from Theorem~\ref{thm:mo2}.
\end{theorem}
Notice that the Morales-Ramis Theorem~\ref{thm:mo2} is a corollary to the above
theorem.

A super-integrable system is integrable in the Liouville sense. Thus necessary
conditions for the super-integrability have to restrict the list~\eqref{eq:tab}
from
Theorem~\ref{thm:mo2}. Our second theorem gives such restrictions. Notice that
these restrictions are imposed on non-trivial eigenvalues. 
\begin{theorem}
\label{thm:we2}
If a Hamiltonian system defined by Hamiltonian~\eqref{eq:ham} with a homogeneous
potential of degree $k\in\Z^\star$ admits $n +l$ , $1< l < n$
functionally
independent  meromorphic first integrals $F_1=H, F_2, \ldots, F_{n+l}$, such
that  $F_1=H, F_2, \ldots, F_{n}$ commute, then each 
$(k,\lambda_i)$ belongs to the list~\eqref{eq:tab} from Theorem~\ref{thm:mo2},
and  moreover 
\begin{itemize}
\item if $\abs{k}\leq 2$, then at least $l$  pairs $(k,\lambda_i)$, where 
$\lambda_i$ is a non-trivial eigenvalue, belong to the following list
\begin{equation}
 \label{eq:tab3}
\begin{array}{crl}
\text{case} &  k & \lambda  \\[0.5em]
\hline
\vbox to 1.3em{}I, & -2 , & \lambda=1-r^2 \\[0.5em]
 II. & -1, & 1 \\[0.5em]
 III. & 1 &    0    \\[0.5em]                         
IV. & 2 & \lambda=r^2\\
\hline
\end{array}
\end{equation}
where $r\in\Q^\star$;
\item if $\abs{k}> 2$, then at least $l$  pairs $(k,\lambda_i)$ where 
$\lambda_i$ is a non-trivial eigenvalue,  belong to items
3--9 of
table~\eqref{eq:tab}.
\end{itemize}
\end{theorem}
\begin{remark}
\label{rem:k2}
Assume that $\vd$  satisfies   $V'(\vd)=\gamma \vd$ with $\gamma\neq 1$ then,
working with straight line solution~\eqref{eq:sls}, we arrive to the same
results. 
However, to apply the above three theorems, we have to make the following
modification. If $\hat\lambda_1, \ldots, \hat\lambda_n$ are eigenvalues of
$V''(\vd)$, then we put $ \lambda_i=\hat\lambda_i/\gamma$, for $i=1,
\ldots n$,  see
\cite{Maciejewski:05::b} for details. This remark is important only for $k=2$,
as for $k\neq2$ we can always assume that  for a straight line
solution~\eqref{eq:sls} Darboux point $\vd$ satisfies $V'(\vd)=\vd$.
\end{remark}
The rest of this paper, except for the last section,  is devoted to present
proofs of the above theorems. 
To this end, in the next section we recall basic facts from the Ziglin 
theory \cite{Ziglin:82::b,Ziglin:83::b} and its differential Galois extension 
developed by A.~Baider,
R.~C.~Churchill, J.~J.~Morales, J.-P.~Ramis, D.~L.~Rod, C.~Sim\'o and
M.~F.~Singer, see
\cite{Churchill:96::b,Morales:99::c,Morales:99::b,Morales:01::b1} and references
therein, which is called the Morales-Ramis theory.  Section~\ref{sec:lio}
contains a derivation of
variational equations and their reduction to an algebraic form. It appears that
these equations are a direct sum of a certain type of hypergeometric
 equations. In section~\ref{sec:gg} we give a detailed analysis of the
differential
Galois group of this type of hypergeometric equation. Obstructions for partial
and super-integrability follow from the fact that the differential Galois
group of variational equations must have an appropriate number of invariants.
This problem, reformulated into  the language of Lie algebra of the differential
Galois group, is analysed in
Section~\ref{sec:pa}. Short proofs of Theorem~\ref{thm:we1} and \ref{thm:we2}
are given in Section~\ref{sec:p}. In the last section we present an application
of our theorems. We analyse three and four body problems on a line and  a radial
potential. To make the paper self-contained, we collect in Appendix several
known facts concerning the differential Galois group of a general second order
equation
with rational coefficients and the Riemann $P$ equation.

\section{Basic facts from the general theory}
\label{sec:basics}
In this section we recall several basic facts from the Ziglin and Morales-Ramis
theory in the setting needed in this paper. For detailed expositions, see e.g.
\cite{Audin:01::c,Audin:02::c,Morales:01::b1,Morales:01::b2,Morales:99::c,
Churchill:96::b}.

Thus let us consider a complex  holomorphic system of differential equations
\begin{equation}
\label{eq:ds}
\Dt \vx = \vv(\vx), \qquad \vx\in U\subset\C^n, \quad t\in\C, 
\end{equation} 
where $U$ is an open and connected subset of $\C^n$.  
The Ziglin and Morales-Ramis theory are based  on the linearization of the
original system
around a particular non-equilibrium solution. Hence, let $\vvarphi(t)$ be a
non-equilibrium solution  of~\eqref{eq:ds}.  Usually it is not a single-valued
function of the complex time $t$. Thus, we associate with $\vvarphi$ a Riemann
surface $\Gamma$ with $t$ as a local coordinate.  The variational equations
along $\Gamma$ have the form
\begin{equation}
\label{eq:vds}
 \dot \vxi = A(t)\vxi, \qquad  A(t)=\frac{\partial
 \vv}{\partial \vx}(\vvarphi(t) ) ,  \qquad \vxi \in T_\Gamma U.
\end{equation}
With these equations we can  associate two groups. The first one, called the
monodromy group  is defined as follows. Let us fix a point $p\in\Gamma$,
and let $\Xi(t)$ be the local fundamental system of solutions of~\eqref{eq:vds}
satisfying $\Xi(t_0)=E$, where $E$ is the identity matrix, and
 $\vvarphi(t_0)=p$. Matrix $\Xi(t)$  is holomorphic for a
sufficiently small disc
$D_\varepsilon(t_0):=\defset{t\in\C}{\abs{t-t_0}<\varepsilon}$.  Then an
analytical continuation
of $\vXi(t)$ along a closed curve $\gamma$ with a base point $p$ gives rise to a
new fundamental system $\widetilde\vXi(t)$  for $t\in D_\varepsilon(t_0)$. 
Solutions of a linear system~\eqref{eq:vds} form a $\C$-linear space, hence
each column in $\widetilde\vXi(t)$ is a linear combination of columns of
$\Xi(t)$.
Thus we have $\widetilde\vXi(t)=\vXi(t) M_\gamma$, for a certain  $
M_\gamma\in\mathrm{GL}(n,\C)$. It can be shown that matrix $M_\gamma$, called
the monodromy matrix,  does not depend on a specific choice of $\gamma$ only on
its homotopy class.  If $\gamma$ is a product of two loops
$\gamma=\gamma_1\cdot\gamma_2$ (first go along loop $\gamma_1$, and then go
along $\gamma_2$), then
\begin{equation*}
 M_\gamma=M_{\gamma_1\cdot\gamma_2}=M_{\gamma_2}M_{\gamma_1},
\end{equation*}
in other words, an analytic continuation of solutions of system~\eqref{eq:vds}
along closed paths with a fixed point $p$, gives an anti-representation of the
first fundamental group $\pi_1(p,\Gamma)$ of $\Gamma$. The image $\cM$ of this
anti-representation is called the monodromy group of equation~\eqref{eq:vds}. In
the above definition a point $p\in \Gamma$ appeared, so we should write $\cM_p$.
However,   if we choose $q\in\Gamma$, $q\neq p$,   then the obtained monodromy
group $\cM_q$ is isomorphic with $\cM_p$. More precisely,  there exists a 
matrix $C\in\mathrm{GL}(n,\C)$ such that every element $A$ of $\cM_q$  is
uniquely given by $C^{-1}BC$, where $B\in\cM_p$. Thus, we do not specify 
the dependence on the base point later.

 To define the differential Galois  group of equation~\eqref{eq:vds} we have to
switch to the algebraic language. We can consider the entries of matrix $A(t)$
in equation~\eqref{eq:vds}  as elements
of  field $K:=\mathscr{M}(\Gamma)$  of functions  meromorphic  on $\Gamma$. 
This
field with the differentiation with respect to $t$ as a derivation is  a
differential field. Only constant functions from $K$ have a vanishing
derivative,
so the subfield of constants of $K$ is $\C$.  
It is obvious that  solutions of~\eqref{eq:vds} are not necessarily elements
of  $K^n$. The fundamental theorem of the differential Galois theory guarantees
that there exists a differential field $L\supset K$ such  that $n$
linearly independent (over $\C$) solutions of~\eqref{eq:vds} are contained in
$L^n$. The smallest
differential extension  $L\supset K$  with this property  is called the
Picard-Vessiot extension of $K$. A group $\mathscr{G}$ of differential
automorphisms of
$L$ which do not change $K$ is called the differential Galois group of
equation~\eqref{eq:vds}. 
It can be shown that $\mathscr{G}$ is a linear algebraic group.  Thus, it is a
union
of a finite number
of disjoint connected components. One of them, containing the identity, is
called
the identity component and is denoted by $\mathscr{G}^{\circ}$.

Let $\vxi=(\xi_1, \ldots,\xi_n)^T\in L^n$ be a solution of
equation~\eqref{eq:vds}, and $g$ an element of its  differential Galois group.
Then,  $g(\vxi):= (g(\xi_1), \ldots,g(\xi_n))^T$ is also its solution. In fact,
by definition $g$ commutes with the time differentiation, so we have
\begin{equation*}
 \Dt g(\vxi)= g(\dot \vxi)= g(A(t)\vxi)=A(t)g(\vxi),
\end{equation*}
as $g$ does not change elements of $K$. Thus, if $\vXi\in \M(n,L)$ is a
fundamental matrix of~\eqref{eq:vds}, i.e., its columns are linearly independent
solutions of~\eqref{eq:vds}, then
$
 g(\vXi)=\vXi M_g
$, where $M_g\in \mathrm{GL}(n,\C)$. In other words, we can look at the
differential Galois group as a matrix group.

It is known that the monodromy group is contained in the differential Galois
group. Moreover,  in a case when equations~\eqref{eq:ds} are Hamiltonian, then
both these groups are subgroups of $\mathrm{Sp}(n,\C)$.

Now we explain why the  monodromy and differential Galois groups of variational
equations  are
important in a study of integrability.  At first, we introduce a few
definitions.
Let us consider 
 a holomorphic function $F$ defined in a
certain connected neighbourhood  of solution $\vvarphi(t)$. In this
neighbourhood we have the expansion
\begin{equation}
F(\vvarphi(t) +\vxi)= F_m(\vxi) + O(\norm{\vxi}^{m+1}),  \qquad F_m\neq 0.
\end{equation}
Then the leading term $f$ of $F$ is the lowest order  term   of the above
expansion
i.e., $f(\vxi):=F_m(\vxi)$. 
Note that  $f(\vxi)$ is a homogeneous polynomial of variables
$\vxi=(\xi_1,\ldots, \xi_n)$ of degree $m$; its coefficients are polynomials in
$\vvarphi(t)$.  If $F$ is a meromorphic function,
then it can be written as $F=P/Q$  for  certain holomorphic functions $P$ and
$Q$.
Then  the
leading term  $f$ of $F$  is defined as $f=p/q$, where $p$ and $q$ are leading
terms of $P$ and $Q$, respectively. In this case $f(\vxi)$ is a homogeneous
rational function of $\vxi$. 

One can prove that if $F$ is a meromorphic (holomorphic) first integral of 
equation~\eqref{eq:ds}, then its leading term $f$ is a rational (polynomial)
first integral of
variational  equations~\eqref{eq:vds}. If system ~\eqref{eq:ds} has
$m\geq 2$
functionally independent meromorphic first integrals $F_1, \ldots, F_m$, then 
their leading terms can be functionally dependent. However,  by
the Ziglin Lemma \cite{Ziglin:82::b,Audin:01::c,Churchill:96::b}, we can find
$m$ polynomials $G_1, \ldots, G_m\in\C[z_1,\ldots, z_m]$ such that leading
terms of $G_i(F_1,\ldots, F_m)$, for $1\leq i \leq m$ are functionally
independent.

Additionally, if $\mathscr{G}\subset \mathrm{GL}(n,\C)$ is the differential
Galois group
of~\eqref{eq:vds}, and $f$ is its rational first integral, then
$f(g(\vxi))= f(\vxi)$ for every $g\in\mathscr{G}$, see
\cite{Audin:01::c,Morales:99::c}.
This means that  $f$ is a rational
invariant of group $\mathscr{G}$. Thus we have a correspondence between
the first
integrals of the system~\eqref{eq:ds} and invariants of  $\mathscr{G}$.
\begin{lemma}
If equation~\eqref{eq:ds} has $k$ functionally independent  first integrals
which
are meromorphic in a connected neighbourhood  a non-equilibrium solution
$\vvarphi(t)$, then the differential Galois group $\mathscr{G}$ of the
variational equations
along  $\vvarphi(t)$,  as well as their monodromy group, have $k$ functionally
independent  rational invariants.
\label{lem:ratinv}
\end{lemma}
As mentioned above, a differential Galois group is a linear algebraic group,
thus, in
particular, it is a Lie group, and one can consider its a Lie algebra. This Lie
algebra
reflects only the properties of  the identity component of the group.
It is easy to show that if a Lie group has an invariant, then also its
Lie algebra has an integral. Let us explain what the last expression means.
Let $\mathfrak{g}\subset \mathrm{GL}(n,\C)$ denote the Lie algebra of
$\mathscr{G}$. Then an
element $Y\in
\mathfrak{g}$ can be considered as a  linear vector field: $\vx\mapsto Y(\vx):=
Y\vx$,
for $\vx\in\C^n$. 
We say that $f\in\C(\vx)$ is an integral of $\mathfrak{g}$, iff $Y(f)(\vx)=\rmd
f(\vx)\cdot Y(\vx)=0$, for all $Y\in \mathfrak{g}$.  
\begin{proposition}
\label{prop:lieint}
If $f_1, \ldots, f_k\in\C(\vx)$ are algebraically independent invariants of an
algebraic  group $\mathscr{G}\subset \mathrm{GL}(n,\C)$, then they are
algebraically independent  first integrals
of the Lie algebra $\mathfrak{g}$ of $\mathscr{G}$.
\end{proposition}
The above facts are the starting points for applications of differential Galois
methods to a study of integrability.

If the considered system is Hamiltonian, then we have additional constrains.
First
of all, the differential Galois group of variational equations is a subgroup of
the symplectic group. Secondly, commutation of first integrals imposed by the
Liouville integrability implies commutation of variational first integrals.
The following lemma plays the crucial role and this is why it was called The Key
Lemma see Lemma~III.3.7 on page 72 in~\cite{Audin:01::c}.
\begin{lemma}
 \label{lem:key}
Assume that Lie algebra $\mathfrak{g}\subset \mathrm{sp}(2k,\C)$ admits $k$
functionally independent and commuting first integrals. Then  $\mathfrak{g}$ is
Abelian.
\end{lemma}
Hence, if $\mathfrak{g}$ in the above lemma is the Lie algebra of a Lie group
$\mathscr{G}$, then the identity component $\mathscr{G}^\circ$ of $\mathscr{G}$
is Abelian.

Using all these facts   Morales and Ramis proved the following
theorem \cite{Morales:99::c,Morales:01::b1}.
\begin{theorem}[Morales-Ramis]
\label{thm:basicG}
Assume that a Hamiltonian system is meromorphically integrable in the
Liouville sense in a neighbourhood of a phase curve
$\Gamma$, and that variational equations along $\Gamma$ are Fuchsian.  Then the
identity component of the differential Galois group of the variational
equations 
 is Abelian.
\end{theorem}
Generally, it is difficult to determine the differential Galois group
of a given system of variational equations when its dimension is greater than
two. This is a reason why, instead of the variational equations, it is
convenient to work with their reduced form called the normal variational
equations. For a general definition of this notion see e.g.
\cite{Morales:99::c}. Here we define  the normal variational
equations for a case when  the considered Hamiltonian system is defined
on $\C^{2n}$ with $\vz=(q_1,p_1,\ldots,q_n,p_n)$   as
canonical coordinates. Let us assume that the system admits a two dimensional
symplectic invariant plane
\begin{equation}
 \Pi:=\defset{\vz\in\C^{2n}}{q_i=p_i=0\mtext{for}i=1,\ldots,n-1}.
\end{equation} 
Thus, if $H$ is the Hamiltonian of the system, then
\begin{equation}
 \pder{H}{q_i}(0,\ldots0,q_n,p_n)=\pder{H}{p_i}(0,\ldots0,q_n,p_n)=0
\mtext{for}i=1,\ldots,n-1.
\end{equation} 
Now, for a particular solution $\vvarphi(t)=(0,\ldots0,q_n(t),p_n(t))$,  the
matrix of the variational equations has a block diagonal form
\begin{equation}
 A(t)=\begin{bmatrix}
       N(t)& 0\\
       B(t)   & T(t)
      \end{bmatrix},
\end{equation} 
where $N(t)$, $B(t)$ and $T(t)$ are $2(n-1)\times 2(n-1)$, $2\times 2(n-1)$
and $2\times 2$ matrices,
respectively. Hence, the variational equations are a product of two
systems 
\begin{equation}
 \Dt \vxi = N(t)\vxi, \quad \vxi\in\C^{2(n-1)} \mtext{and} \Dt \veta =
B(t)\vxi + T(t)\veta, \quad \veta\in\C^{2}.
\end{equation} 
The first of them is called the normal
variational equations.  

It can be shown that if the Hamiltonian system possesses a first integral $F$, 
then the normal variational equations also have a first integral which is an
invariant of
their differential Galois group
$\mathscr{G}_{\mathrm{N}}\subset\mathrm{Sp}(2(n-1),\C)$. Moreover, if
$F_1$ and $F_2$ are commuting first integrals functionally independent together
with $H$, then we can assume that the corresponding first integrals $f_1$ and
$f_2$ of the normal variational equations are independent and commuting, see
\cite{Churchill:96::b,Morales:99::c}. These facts imply that the statement of
Theorem~\ref{thm:basicG} remains valid if in its formulation  the normal
variational equations are used instead of the variational equations. 

\section{Necessary conditions for Liouville integrability}
\label{sec:lio}

Let $G(k,\lambda)$ denote the differential Galois group of equation
\begin{equation}
\label{eq:solk}
 \ddot y = -\lambda \varphi(t)^{k-2} y.
\end{equation}
It is a subgroup of $\mathrm{SL}(2,\C)\simeq\mathrm{Sp}(2,\C)$. 

 It is clear that the differential Galois group $G$, of
equations~\eqref{eq:varts} is a direct product
\begin{equation}
 \label{eq:dirsumg}
G=G(k,\lambda_1)\times \cdots \times G(k,\lambda_n)\subset 
\mathrm{Sp}(2n,\C). 
\end{equation} 
Hence, $G^\circ$ is Abelian if and only if  groups $G(k,\lambda_i)^\circ$ are
Abelian, for $i=1,\ldots,n$.  It follows that we know for which values of
$k$ and $\lambda$ the identity component  $G(k,\lambda)^\circ$ of the
differential Galois group  $G(k,\lambda)$ of equation~\eqref{eq:solk} is
Abelian.

To solve this problem we introduce a new independent variable in
equation~\eqref{eq:solk}, as it was proposed  in~\cite{Yoshida:87::a},  namely,
assuming that $k\neq 0$ and $e\neq 0$ we put
\begin{equation}
\label{eq:yo}
 t\rightarrow z:=\frac{1}{ek} \varphi(t)^k.
\end{equation}
Then, equation ~\eqref{eq:solk} is transformed to the following one
\begin{equation}
\label{eq:hyps}
z(1-z)y'' + \left(\frac{k-1}{k} - \frac{3k-2}{2k}z\right)y' +
\frac{\lambda}{2k} y=0 ,
\end{equation} 
where prime denotes the differentiation with respect to $z$. It is the Gauss
hypergeometric equation
\begin{equation}
z(1-z)y'' +[c-(a+b+1)z]y' -ab y=0,
\label{eq:hyp}
\end{equation}
with parameters
\begin{equation}
a+b=\frac{k-2}{2k},\qquad ab=-\frac{\lambda}{2k},\qquad c=1-\frac{1}{k}.
\label{eq:abc}
\end{equation}
The differences of exponents at $z=0,1$, and
$\infty$  for
equation~\eqref{eq:hyp} are
\begin{equation}
\label{eq:expdiff}
\rho=1-c=\frac{1}{k}, \qquad \sigma=c-a-b=\frac{1}{2} ,\qquad \tau=a-b=
\frac{1}{2k}\sqrt{(k-2)^2+8k\lambda},
\end{equation}
respectively.

Let $\widehat G(k,\lambda)$ denote the differential Galois group of
equation~\eqref{eq:hyps}.  Notice that $\widehat G(k,\lambda)$ is a subgroup of
$\mathrm{GL}(2,\C)$, and is different from $G(k,\lambda)$.  However, it can be
shown, see~\cite{Morales:01::a,Morales:99::c},  that $G(k,\lambda)^\circ$ and
$\widehat G(k,\lambda)^\circ$ are isomorphic.

Now, the change of the independent variable~\eqref{eq:yo}, transforms the
variational equations~\eqref{eq:varts} into a direct product of hypergeometric
equations
\begin{equation}
\label{eq:veh}
z(1-z)y_i'' + \left(\frac{k-1}{k} - \frac{3k-2}{2k}z\right)y_i' +
\frac{\lambda_i}{2k} y_i=0 , \qquad  1\leq i\leq n,
\end{equation} 
whose differential Galois group $\widehat G$ is a direct product
\begin{equation*}
 \widehat G= \widehat G(k,\lambda_1)\times\cdots\times\widehat G(k,\lambda_n ).
\end{equation*}
A necessary condition for the integrability is now following: all groups $
\widehat G(k,\lambda_i)^\circ$ have to be Abelian, and thus solvable. Exactly
this reasoning was used in the proof of Theorem~\ref{thm:mo2} given
in~\cite{Morales:99::c,Morales:01::a}.

\section{Group $ G(k,\lambda)^\circ$}
\label{sec:gg}
From the previous section it follows that it is important to know precisely  the
identity component of the differential Galois group of hypergeometric
equation~\eqref{eq:hyp} with parameters $a$, $b$ and $c$ given
by~\eqref{eq:abc}. This is the aim of this section.

 As we have already mentioned the differential Galois group of
~\eqref{eq:hyp} is not a subgroup of $\mathrm{SL}(2,\C)$.  It causes some
technical problems.  To avoid them, we transform equation~\eqref{eq:hyp} to the
normal form putting
\begin{equation}
\label{eq:ton}
w= y \exp\int p\, \rmd z, \qquad p:=\frac{c-(a+b+1)z}{z(1-z)}.
\end{equation}
Then we obtain
\begin{equation}
 \label{eq:nhyp}
w'' = \frac{\rho^2 -1+z(1-\rho^2-\tau^2+\sigma^2) +z^2(\tau^2-1)}{4z^2(z-1)^2}
w.
\end{equation}
For this equation exponents at $0$, $1$ and at the  infinity are
\begin{equation}
\label{eq:exphypn}
 \left\{  \frac{1}{2}(1-\rho),  \frac{1}{2}(1+\rho) \right\}, \quad  \left\{ 
\frac{1}{2}(1-\sigma),  \frac{1}{2}(1+\sigma) \right\}, \quad  \left\{ -
\frac{1}{2}(1-\tau), - \frac{1}{2}(1+\tau) \right\},
\end{equation} 
respectively. Its monodromy and differential Galois groups are now  subgroups of
$\mathrm{SL}(2,\C)$.  It is important to remark here that the identity
components of the differential Galois groups of~\eqref{eq:hyp}
and~\eqref{eq:nhyp} are the same. 
Notice also that the differences of exponents at singular points were
unchanged.

Assuming that  $ \rho$,  $\sigma$
and $\tau$ are defined by~\eqref{eq:expdiff}, we denote by
$\mathscr{G}(k,\lambda)$
 the differential Galois group of equation~\eqref{eq:nhyp}. In what follows we
describe properties of $\mathscr{G}(k,\lambda)^\circ$, but, as we explained,
groups $\mathscr{G}(k,\lambda)^\circ$, $\widehat G(k,\lambda)^\circ$ and $
G(k,\lambda)^\circ$ are isomorphic, so, as a result, we obtain a
characterisation of
$G(k,\lambda)^\circ$.

At first, we recall the following fact which explains the origin of
table~\eqref{eq:tab} given in Theorem~\ref{thm:mo2}.
\begin{proposition}
 \label{prop:sol}
Group $\mathscr{G}(k,\lambda)^\circ$ is solvable if and only if $(k,\lambda)$
belongs to an item in table~\eqref{eq:tab}.
\end{proposition}
\begin{proof}
 Equation~\eqref{eq:nhyp} is the Riemann $P$ equation. The  Kimura
theorem~\ref{thm:kimura} gives  necessary and sufficient conditions  for the
solvability of the identity component of its differential Galois group.
Table~\eqref{eq:tab} is just a specification of these conditions for $ \rho$, 
$\sigma$
and $\tau$ given by~\eqref{eq:expdiff}. 
\end{proof}
A necessary condition for the integrability is that
$\mathscr{G}(k,\lambda)^\circ$ is Abelian.  As not all
solvable groups are Abelian, one can think that conditions of
Theorem~\ref{thm:mo2} can be sharpened.  We show that it is not like that,
i.e., we prove  that if $\mathscr{G}(k,\lambda)^\circ$ is solvable, then 
it is Abelian. Suppose that $\mathscr{G}(k,\lambda)^\circ$ is solvable but not
Abelian.  Then,
as it is explained in Appendix, there is only one
possibility: $\mathscr{G}(k,\lambda)= \mathscr{G}(k,\lambda)^\circ=\mathscr{T}$,
where
$\mathscr{T}$ is the triangular subgroup of $\mathrm{SL}(2,\C)$. So, such case
can
appear only if the considered equation is reducible. Let us recall, see
Appendix, that equation~\eqref{eq:nhyp} is reducible iff it has a solution 
$w = \exp[\int \omega]$ where $\omega\in\C(z)$.
\begin{proposition}
 \label{prop:red}
Equation~\eqref{eq:nhyp} is reducible if and only if $\lambda= p+kp(p-1)/2$ for
some $p\in\Z$.
\end{proposition}
\begin{proof}
 To proof this lemma it is enough to check directly  one of equivalent
conditions given in Lemma~\ref{lem:redrie}. 
\end{proof}
If equation~\eqref{eq:nhyp} is reducible,  then respective exponents at singular
points $0$, $1$ and infinity are following
\begin{equation}
\label{eq:expnhr}
 \left\{ \frac{1}{2} -\frac{1}{2k}, \frac{1}{2} + \frac{1}{2k} \right\},
\qquad  \left\{\frac{1}{4}, \frac{3}{4}\right\}, \qquad
\left\{   -\frac{2+k(l+2) }{4k},  \frac{2+k(l-2) }{4k}  \right\},
\end{equation}
where $l$ is an odd integer.

Now, we can show that if equation~\eqref{eq:nhyp} is reducible, then the
identity component of its differential Galois group
$\mathscr{G}(k,\lambda)^\circ$ is a proper subgroup of the triangular group
$\mathscr{T}$, and thus it is Abelian.
\begin{lemma}
 \label{lem:solab}
Assume that equation~\eqref{eq:nhyp} is reducible. Then its differential Galois
group $\mathscr{G}(k,\lambda)$  is a proper subgroup of the triangular group.
\end{lemma}
 \begin{proof}
  The difference of exponents for singular point $z=1$ is  $1/2$. Thus,
from Lemma~\ref{lem:iwa1},  it follows that if
equation~\eqref{eq:nhyp} is reducible, 
then it possesses a solution of the form:
\begin{equation*}
 w =z^r(1-z)^s h(z), 
\end{equation*}
where $h(z)$ is a polynomial, and $r$ is an exponent at $z=0$,  and $s$ in an
exponent at $z=1$. As $r$ and $s$ are rational,
there exists $j\in\N$ such that $w^j\in\C(z)$. Now, by
Lemma~\ref{lem:algc1},  $\mathscr{G}(k,\lambda)$ is either a proper subgroup of
the diagonal group, or a proper subgroup of the triangular group.
 \end{proof}
For our further analysis it is important to know the dimension of  
$\mathscr{G}(k,\lambda)^\circ$ in a case when $\mathscr{G}(k,\lambda)$ is
reducible. By the Lemma~\ref{lem:algc1}           , either
$\mathscr{G}(k,\lambda)$ is a
finite cyclic group, and then $\mathscr{G}(k,\lambda)^\circ=\{E\}$, or
$\mathscr{G}(k,\lambda)$ is a proper subgroup of the triangular group, and then
\begin{equation}
 \label{eq:G0}
\mathscr{G}(k,\lambda)^\circ=\mathscr{T}_1 := \defset{
\begin{bmatrix}
 1 & c\\
0 & 1
\end{bmatrix}
}{c\in\C}.
\end{equation} 
\begin{proposition}
 \label{prop:diag}
Assume that $\mathscr{G}(k,\lambda)$ is diagonal. Then $k\in\{\pm 1,\pm2\}$.
\end{proposition}
\begin{proof}
If $\mathscr{G}(k,\lambda)$ is diagonal, then the monodromy group of
equation~\eqref{eq:nhyp} is diagonal. This last group is generated by  two
elements $M_0$, $M_1\in \mathrm{SL}(2,\C)$ which can be assumed diagonal. Then
from Lemma~\ref{lem:iwa1} it follows that at least one of matrices $M_0$, $M_1$
or $M_0M_1$ is  $\pm E$. The eigenvalues of $M_0$ are $\exp[2\pi\rmi r_{1,2}]$,
where $r_{1,2}$ are exponents at $z=0$ for equation~\eqref{eq:nhyp}, i.e.
\begin{equation*}
 r_{1,2}=\frac{1}{2}\left(1\pm\frac{1}{k}\right).
\end{equation*}
Hence, if  $M_0= E$, then $k=\pm 1$, and it is impossible that $M_0=-E$. Similar
arguments show that $M_1\neq \pm E$, and if  $M_0M_1=\pm E$, then $k=\pm 2$.
\end{proof}
If a local solution near a singular points contains a logarithm, then the
monodromy group contains the following element
\begin{equation*}
 M=\begin{bmatrix}
    1 & 2\pi\rmi\\
    0 & 1
   \end{bmatrix},
\end{equation*}
and, moreover it can be shown that such element belongs to the identity
component of the differential Galois group of considered equation. Hence we can
use this fact for checking whether
$\mathscr{G}(k,\lambda)^\circ=\mathscr{T}_1$. 
\begin{proposition}
 Assume that $k=1$ and that equation~\eqref{eq:nhyp} is reducible. Then
singular point $z=0$ is logarithmic except for the case $\lambda=0$.
\end{proposition}
\begin{proof}
 We apply Lemma~\ref{lem:iwa3} from Appendix, and we use notation introduced
just
before it.
For $k=1$ exponents at $z=0$ are $\rho_1=1$ and $\rho_2=0$,
so $m:=\rho_1-\rho_2=1$, and $\langle m\rangle=\{1\}$.  Thus, by
Lemma~\ref{lem:iwa3}, singularity $z=0$ is logarithmic if an only if
 for arbitrary exponents  $s$ and $t$ at $z=1$ and $z=\infty$, respectively, we
have $1+s
+t\neq 1$. Using \eqref{eq:expnhr} we obtain
\begin{equation}
 1\pm \frac{p}{2}\neq 1 \mtext{and} \frac{3}{2}+\frac{p}{2}\neq 1,
\mtext{for}p\in\Z.
\end{equation} 
This system of inequalities is satisfied for $p\in\Z\setminus\{-1,0\}$. For
$p=0$  and $p=-1$ we have $\lambda=0$. This finishes the proof.
\end{proof}
In a similar way one can show the following.
\begin{proposition}
 Assume that $k=-1$, and that equation~\eqref{eq:nhyp} is reducible. Then
singular point $z=0$ is logarithmic except for the case 
$\lambda=1$.
\end{proposition}
For $k=\pm 2$, the singular point at infinity can be logarithmic. 
\begin{proposition}
 Assume that $k=2$ and that equation~\eqref{eq:nhyp} is reducible. Then
singular point $z=\infty$ is  logarithmic if and only if 
$\lambda=0$.
\end{proposition}
\begin{proof}
 Under the given assumptions,  $\lambda=p^2$, exponents at infinity are
$\tau_1=(p-1)/2$ and $\tau_2=-(p+1)/2$,   where $p\in\Z$.  Thus,
$m:=\tau_1-\tau_2=p$.  If $p=0$, then the singularity is logarithmic. Assume
that $p>0$.  By
Lemma~\ref{lem:iwa3}, singularity $z=\infty$ is logarithmic if an only if
 for arbitrary exponents  $r$ and $s$ at $z=0$ and $z=1$, respectively, we
have
\begin{equation*}
 \frac{1}{2}(p-1) +r + s\not\in\langle m \rangle.
\end{equation*}
Using \eqref{eq:expnhr} we obtain the following condition: none of the three
numbers
\begin{equation*}
 \frac{1}{2}p, \qquad \frac{1}{2}(p+1),  \qquad \frac{1}{2}(p+2),
\end{equation*}
belongs to $\langle m \rangle=\{1, \ldots, p\}$. This is not true, as either the
first or the second belongs to $\langle m \rangle$. Similar arguments work for
$p<0$, and this finishes the proof.
\end{proof}
\begin{proposition}
 Assume that $k=-2$ and that equation~\eqref{eq:nhyp} is reducible. Then
singular point $z=\infty$ is  logarithmic if and only if 
$\lambda=1$. 
\end{proposition}
Let us summarise our analysis.
\begin{corollary}
 \label{col:red}
Assume that equation~\eqref{eq:nhyp} is reducible. Then
$\mathscr{G}(k,\lambda)^\circ=\mathscr{T}_1$ except for the following cases:
\begin{enumerate}
 \item $k=-2$ and  $\lambda=1-p^2$,   $p\in\Z^\star$,
\item $k=-1$ and $\lambda=1$,
\item $k=1$ and $\lambda=0$,
\item $k=2$ and $\lambda=p^2$, $p\in\Z^\star$,
\end{enumerate}
when  $\mathscr{G}(k,\lambda)^\circ=\{E\}$.
\end{corollary}
Let us assume now that $\mathscr{G}(k,\lambda)^\circ$ is not reducible but
solvable. Then we have two possibilities. Either $\mathscr{G}(k,\lambda)$ is
primitive and finite, and then $\mathscr{G}(k,\lambda)^\circ=\{E\}$, or
$\mathscr{G}(k,\lambda)$ is a subgroup of $\mathscr{DP}$  group, see Appendix.
\begin{proposition}
 \label{prop:ch}
Assume that equation~\eqref{eq:nhyp} is not reducible. Then
$\mathscr{G}(k,\lambda)$ is a subgroup of $\mathscr{DP}$ group only if and only
if either:
\begin{enumerate}
\item $k=-2$; in this case $\mathscr{G}(k,\lambda)^\circ=\{E\}$ if and only if 
      $\lambda=1-r^2$ for some  $r\in\Q\setminus\Z$, and
$\mathscr{G}(k,\lambda)^\circ=\mathscr{D}$ otherwise, or
\item $k=2$; in this case $\mathscr{G}(k,\lambda)^\circ=\{E\}$ if and only if 
      $\lambda=r^2$ for some  $r\in\Q\setminus\Z$, and
$\mathscr{G}(k,\lambda)^\circ=\mathscr{D}$ otherwise, or
\item $\abs{k}>2$ and
 \begin{equation*}
    \lambda=   \dfrac 1 {2}\left(\dfrac {k-1} {k}+p(p+1)k\right), \qquad p\in\Z,
      \end{equation*}
and in this case $\mathscr{G}(k,\lambda)$ is finite, so 
$\mathscr{G}(k,\lambda)^\circ=\{E\}$. 
\end{enumerate}
\end{proposition}
\begin{proof}
  We apply Lemma~\ref{lem:dp} from Appendix. A necessary condition for
$\mathscr{G}(k,\lambda)$ to
be a subgroup of $\mathscr{DP}$ group is following: at least two differences of
exponents are half integers. As the difference of exponents at $z=1$ is
$\sigma=1/2$,  we have two possibilities: either $k=\pm 2$ and then $\rho=\pm
1/2$, or 
\begin{equation}
\label{eq:c2}
 \lambda=   \dfrac 1 {2}\left(\dfrac {k-1} {k}+p(p+1)k\right),
\end{equation}
for some $p\in\Z$. Moreover, if $\mathscr{G}(k,\lambda)$ is a subgroup of
$\mathscr{DP}$ group,  then it is a finite group if and only if, at two singular
points, the differences of exponents are half integers and exponents at the
remaining point are rational, otherwise  $\mathscr{G}(k,\lambda)=\mathscr{DP}$.
Hence, under the assumption of our lemma, for $\abs{k}>2$, 
group $\mathscr{G}(k,\lambda)$ is a subgroup of $\mathscr{DP}$ group iff
$\lambda$ is given by \eqref{eq:c2}. 
But, for these values of $\lambda$ all exponents are rational,
and this implies that the group is finite. This proves case $3$. 

For $k=\pm 1$ and $\lambda$ given \eqref{eq:c2} equation~\eqref{eq:nhyp} is
reducible, but we assumed that it is not reducible, so this case is excluded.

Let $k=-2$. Then exponents at infinity are rational if $\tau$ is rational, see
\eqref{eq:exphypn}.  From~\eqref{eq:expdiff} we have
\begin{equation*}
 \tau=-\sqrt{1-\lambda}.
\end{equation*}
 Thus
$\mathscr{G}(-2,\lambda)$ is a finite subgroup of $\mathscr{DP}$ group iff
$\lambda= 1-r^2$ for a rational $r$. However, if $r\in\Z$, then
equation~\eqref{eq:nhyp} is reducible. This proves case
$1$. 

For $k=2$ we have $\tau=\sqrt{\lambda}$, so $\mathscr{G}(2,\lambda)$ is a
finite subgroup of $\mathscr{DP}$ group iff
$\lambda= r^2$ for a rational $r$ but if $r$ is an integer, then
equation~\eqref{eq:nhyp} is reducible, thus we have to exclude these values.
\end{proof}
We summarise our analysis in the following corollary.
\begin{corollary}
\label{cor:0}
Assume that the identity component $\mathscr{G}(k,\lambda)^\circ$ of the
differential Galois group of equation~\eqref{eq:nhyp} is solvable. Then
$\mathscr{G}(k,\lambda)^\circ$ is Abelian. Moreover,
$\mathscr{G}(k,\lambda)^\circ=\{E\}$ if and only if either 
\begin{enumerate}
 \item $\abs{k}>2$ and $(k,\lambda)$ belongs to an item $3$--$9$ of
table~\eqref{eq:tab}, or
\item $\abs{k}\leq 2$ and $(k,\lambda)$ belongs to an item of the following
table
\begin{equation}
 \label{eq:tab3d}
\begin{array}{crl}
\text{case} &  k & \lambda  \\[0.5em]
\hline
I, & -2 , & \lambda=1-r^2 \\[0.5em]
 II. & -1, & 1 \\[0.5em]
 III. & 1 &    0    \\[0.5em]                         
IV. & 2 & \lambda=r^2\\
\hline
\end{array}
\end{equation}
where $r\in\Q^\star$.
\end{enumerate}

\end{corollary}
\section{Certain Poisson algebra}
\label{sec:pa}
As was mentioned for a Hamiltonian system, the differential Galois group
$\mathscr{G}$ of variational equations along a particular solution  is a
subgroup of the
symplectic group $\mathrm{Sp}(2n,\C)$,  thus the Lie algebra $\mathfrak{g}$ is a
Lie subalgebra of $ \mathrm{sp}(2n,\C)$.    The necessary conditions for the
integrability in the Liouville sense from Theorem~\ref{thm:basicG}, are
expressed in terms of the identity component of $\mathscr{G}$. The properties of
this component are encoded in the Lie algebra $\mathfrak{g}$ of $\mathscr{G}$.
To find  the necessary conditions for partial and super-integrability, we  have
to
characterise  Lie algebras  $\mathfrak{g}$ which admit a certain number of first
integrals. And this is the main goal of this section.  Here we follow the ideas
and
methods introduced  in~\cite{mp:07::j}.

An element $Y$ of  Lie algebra
$ \mathrm{sp}(2n,\C)$, considered as a linear vector field, is a Hamiltonian
vector field  given by a global Hamiltonian function $H:\C^{2n}\rightarrow \C$,
which is a  degree $2$ homogeneous
polynomial of  $2n$ variables $(\vx,\vy):=(x_1,\ldots,x_n,y_1,\ldots,y_n)$,
i.e. $H\in
\C_2[\vx,\vy]$.   In this way we identify 
$\mathrm{sp}(2n,\C)$  with a $\C$-linear vector space
$\C_2[\vx,\vy]$ with
the canonical Poisson bracket as the Lie bracket.  Thus, for a Lie algebra
$\mathfrak{g}\subset
\mathrm{sp}(2n,\C)\simeq \C_2[\vx,\vy]$, a rational function $f\in\C[\vx,\vy]$
is a first integral of $\mathfrak{g}$, iff $\{H,f\}=0$, for all $H\in
\mathfrak{g}$. A field of rational first integrals of  $\mathfrak{g}$ we denote
by $\C(\vx,\vy)^{\mathfrak{g}}$.

Now, we consider  the case when $ \mathfrak{g}$ is a Lie subalgebra
of
$\mathrm{sp}(2,\C)$. It is easy to show that Lie algebra $\mathrm{sp}(2,\C)$
does not admit any non-constant first integral.
\begin{proposition}
\label{prop:sp2}
A rational function  $f\in\C(x,y)$ is a first integral of $\mathrm{sp}(2,\C)$,
iff $f\in\C$.
\end{proposition}
\begin{proof}
Let $f\in \C(x,y)$ be a first integral of $\mathrm{sp}(2,\C)\simeq
\C_2[x,y]$. Thus, $\{f,H\}=0$, for each $H\in\C_2[x,y]$. Let us take $H=x^2$.
Then,
\begin{equation*}
\{f,H\}= -2x\pder{f}{y}=0, 
\end{equation*}
and this shows that $f$ does not depend on $y$, i.e., $f\in\C(x)$.  Taking
$H=y^2$, we show that
$f$ does not depend on $x$. Hence $f\in\C$. 
\end{proof}
The above proposition shows that only  proper subalgebras of $\mathrm{sp}(2,\C)$
can have non-constant first integrals.
\begin{proposition}
\label{prop:dim1}
If $\mathfrak{g}$ is a Lie subalgebra of $\mathrm{sp}(2,\C)$ and $\dim_{\C}
\mathfrak{g}>0$, then the number of algebraically independent rational first
integrals
of $\mathfrak{g}$ is not greater than one.
\end{proposition}
\begin{proof}
As $\dim_{\C} \mathfrak{g}>0$, there exists a non-zero $H\in
\mathrm{sp}(2,\C)\simeq
\C_2[x,y]$. The number of  rational algebraically independent  first integrals
of
a  non-zero linear Hamiltonian vector field $X_H$  in $\C^2$ is  at most one.
\end{proof}
\begin{proposition}
\label{prop:dim2}
If $\mathfrak{g}$ is a Lie subalgebra of $\mathrm{sp}(2,\C)$ and $\dim_{\C}
\mathfrak{g}=2$, then  $\C(x,y)^{\mathfrak{g}}=\C$.
\end{proposition}
\begin{proof}
 All two dimensional Lie algebras are solvable so  $ \mathfrak{g}$ is solvable.
Thus
a connected Lie group
 $G\subset \mathrm{sp}(2,\C)$ with  Lie algebra  $\mathfrak{g}$ is solvable.  By
the
Lie-Kolchin theorem $G$ is conjugate  to the triangular group
\begin{equation}
\mathscr{T}:=\defset{ \begin{bmatrix}
                      a& b \\
                      0 &a^{-1}
                      \end{bmatrix}}{ a\in\C^\star, \quad b\in\C}.
\end{equation}
The Lie algebra $\mathfrak{t}$ of  $\mathscr{T}$ is isomorphic to
$\mathfrak{g}$, and is generated by two
elements
\begin{equation}
h_1 = \begin{bmatrix}
       1 & 0\\
       0 & -1
      \end{bmatrix}, \qquad
  h_2 = \begin{bmatrix}
       0 & 1\\
       0 & 0
      \end{bmatrix}.
\end{equation}
 Let $H_1$ and $H_2$ be Hamiltonian functions from $\C_2[x,y]$ such that linear
vector fields $X_{H_1}$ and $X_{H_2}$ have matrices $h_1$ and $h_2$,
respectively. It is easy to check that
\begin{equation}
H_1 = xy, \qquad H_2= \frac{1}{2}y^2.
\end{equation}
We show that  $\C(x,y)^{\mathfrak{t}}=\C$.
Assume that there exists $f\in\C(x,y)^{\mathfrak{t}}\setminus\C$.   Hence
$\{f,H_i\}=0$ for $i=1,2$. But
\begin{equation*}
\{f,H_2\}= y\pder{f}{x} = 0,
\end{equation*}
so, $ f\in\C(y)$. However,  for $ f\in\C(y)$, we have 
\begin{equation*}
\{f,H_1\}= -y\pder{f}{y} = 0,
\end{equation*}
and this implies that $f\in\C$. A contradiction with assumption that $f$ is not
a constant shows that $\C(x,y)^{\mathfrak{t}}=\C$.  Moreover, as Lie algebras
 $\mathfrak{t}$ and  $\mathfrak{g}$ are isomorphic, we have also
$\C(x,y)^{\mathfrak{g}}=\C$.
\end{proof}

Now, we consider a case adopted for a variational equation of the
form~\eqref{eq:varts}. For such equations the differential Galois group
$\mathscr{G}$  is a direct product
\begin{equation}
\mathscr{G}= \mathscr{G}_1 \times \cdots \times \mathscr{G}_n, 
\end{equation}
where $\mathscr{G}_i$ is an algebraic subgroup of $\mathrm{Sp}(2,\C)$, for
$i=1,\ldots,n$.  Hence, the Lie algebra $\mathfrak{g}$ of $\mathscr{G}$ is
also a direct sum
\begin{equation}
\mathfrak{g} = \mathfrak{g}_1\oplus\cdots\oplus\mathfrak{g}_n,
\end{equation} 
where $\mathfrak{g}_i$ is a Lie subalgebra of $\mathrm{sp}(2,\C)$, for
$i=1,\ldots,n$. Let us denote by $\mathfrak{s}_n$ the Lie algebra which is the
direct sum of $n$ copies of  $\mathrm{sp}(2,\C)$
\begin{equation}
\label{eq:s}
\mathfrak{s}_n:=
\underbrace{\mathrm{sp}(2,\C)\oplus\cdots\oplus\mathrm{sp}(2,\C)}_{n-\text{times
}},
\end{equation}
 and by 
$\pi_i: \mathfrak{s}_n\rightarrow \mathrm{sp}(2,\C)$, the projection onto the
$i$-th component of $\mathfrak{s}_n$, for $i=1,\ldots, n$. If we make
identification $\mathrm{sp}(2n,\C)\simeq\C_2[\vx,\vy]$, then $\mathfrak{s}_n$ is
viewed as
\begin{equation}
 \mathfrak{s}_n=\bigoplus_{i=1}^n\C_2[x_i,y_i].
\end{equation} 

\begin{lemma}
\label{lem:one}
 Let $\mathfrak{g}$ be a Lie subalgebra of $\mathfrak{s}_n$,   and
$\mathfrak{g}_i=\pi_i(\mathfrak{g})$ for $i=1,\ldots,n$. Assume that
$f\in\C(\vx,\vy)$ is a non-constant rational first integral of
$\mathfrak{g}$, and that there exists $1\leq j\leq n$, such that
$\mathfrak{g}_j$ is not Abelian, Then $f$ does not depend on $x_j$ and $y_j$.
\end{lemma}
\begin{proof}
 As  $\mathfrak{g}_j$ is not Abelian its dimension is greater
than one.
Let us consider the case $\dim_{\C}\mathfrak{g}_j=3$. Then
$\mathfrak{g}_j=\mathrm{sp}(2,\C)$, and we proceed as in the proof
of~Proposition~\ref{prop:sp2}.
First integral $f$ of  $\mathfrak{g}$ is, in particular, a first integral of
$\mathfrak{g}_j$. Hence, for each $H\in\mathfrak{g}_j\simeq\C_2[x_j,y_j]$ we
have $\{H,f\}=0$. For $H=x_j^2$, we obtain
\begin{equation*}
 0=\{H,f\}=-2x_j\pder{f}{y_j},
\end{equation*}
so $f$ does not depend on $y_j$. Taking $H=y_j^2$ we show that $f$ does not
depend on $x_j$.

If $\dim_{\C}\mathfrak{g}_j=2$, then  we proceed in a similar way using
arguments from the proof of Proposition~\ref{prop:dim2}.
\end{proof}
From the above lemma we have the following consequences.
\begin{corollary}
\label{cor:tri}
 Let $\mathfrak{g}$ be a Lie subalgebra of $\mathfrak{s}_n$,   and
$\mathfrak{g}_i=\pi_i(\mathfrak{g})$ for $i=1,\ldots,n$. If $\mathfrak{g}_i$ is
not Abelian for  $i=1,\ldots,n$, then $\C(\vx,\vy)^{\mathfrak{g}}=\C$.
\end{corollary}
\begin{corollary}
\label{cor:gi}
 Let $\mathfrak{g}$ be a Lie subalgebra of $\mathfrak{s}_n$,   and
$\mathfrak{g}_i=\pi_i(\mathfrak{g})$ for $i=1,\ldots,n$. If
$f\in\C(\vx,\vy)\setminus\C$ is a first integral of $\mathfrak{g}$, then there
exists $1\leq j \leq n$ such that $\mathfrak{g}_j$ is Abelian.
\end{corollary}
Now, we consider a case when $\mathfrak{g}$ admits more than one independent
first integral. 
\begin{lemma}
\label{lem:bas}
Let $\mathfrak{g}$ be a Lie subalgebra of $\mathfrak{s}_n$,   and
$\mathfrak{g}_i=\pi_i(\mathfrak{g})$ for $i=1,\ldots,n$. If $\mathfrak{g}$
admits two algebraically independent and commuting first integrals
$f,g\in\C(\vx,\vy)$, then there exist $1\leq i <j \leq n$, such that
$\mathfrak{g}_i$ and $\mathfrak{g}_j$ are Abelian.
\end{lemma}
\begin{proof}
Neither $f$ nor $g$ is a constant. Thus, by Corollary~\ref{cor:gi}, there
exists $1\leq i \leq n$ such that $\mathfrak{g}_i$ is Abelian. Without loss of
 generality we can assume that $\mathfrak{g}_1$ is Abelian. Suppose that
$\mathfrak{g}_j$ for $ 2\leq j\leq n$ are not Abelian. Then, from
Lemma~\ref{lem:one}, it follows that $f$ and $g$ do not depend on $(x_j,y_j)$
for  $ 2\leq j\leq n$, and hence $f,g\in\C(x_1,y_1)$. But $f$ and $g$ commute,
thus
\begin{equation*}
 0=\{f,g\}=\pder{f}{x_1}\pder{g}{y_1}-\pder{f}{y_1}\pder{g}{x_1}=\frac{\partial
(f,g)}{\partial(x_1,y_1) },
\end{equation*}
so $f$ and $g$ are functionally, and thus algebraically dependent. A
contradiction finishes the proof.
\end{proof}
To prove the next lemma we need the following well known fact, see
e.g. Proposition~3.7 on page 63 in~\cite{Morales:99::c}.
\begin{proposition}
\label{prop:maxi}
 Consider  $\C^{2m}$ as a linear symplectic space with the canonical
coordinates $(\vq,\vp)=(q_1,\ldots,q_m,p_1,\ldots,p_m)$. If $f_1, \ldots, f_l\in
\C(\vq,\vp)$ are algebraically independent and commuting, then $l\leq m$.
\end{proposition}
\begin{lemma}
\label{lem:bask}
Let $\mathfrak{g}$ be a Lie subalgebra of $\mathfrak{s}_n$,   and
$\mathfrak{g}_i=\pi_i(\mathfrak{g})$ for $i=1,\ldots,n$. If $\mathfrak{g}$
admits  first integrals $f_1, \ldots, f_p\in\C(\vx,\vy)$, where  $1<p\leq n$, 
which
are algebraically independent and commuting, then there exist $1\leq i_1 <
\ldots <i_p\leq n$, such that
$\mathfrak{g}_{i_1}, \ldots \mathfrak{g}_{i_1}$ are Abelian.
\end{lemma}
\begin{proof}
 We prove this lemma by induction with respect to $p$. We have already proved
this
lemma
 for $p=2$. Assume that this lemma is valid for $p=j$, where  $2<j<n$. We
show that it is valid for $p=j+1$.

We have $j+1$ commuting and independent first integrals $f_1, \ldots, f_{j+1}$.
From the inductive assumption it follows that among all $\mathfrak{g}_i$ at
least
$j$ are Abelian. We can assume that $\mathfrak{g}_1, \ldots, \mathfrak{g}_j$ are
Abelian. We have
to show that there exits  $ j<l\leq n$ such that $\mathfrak{g}_l$ is Abelian. We
prove it
by contraction. Thus assume that $\mathfrak{g}_{j+1}, \ldots,
\mathfrak{g}_n$ are not Abelian. Then, from
Lemma~\ref{lem:one} it follows that integrals  $f_1, \ldots, f_{j+1}$ do not
depend on $(x_i,y_i)$ for $i=j+1, \ldots, n$. Thus, $ f_1, \ldots,
f_{j+1}\in\C(x_1, \ldots, x_j, y_1, \ldots, y_j)$, and we have a contradiction
with~Proposition~\ref{prop:maxi}. This finishes the proof. 
\end{proof}
\begin{corollary}
\label{cor:bask}
Let $\mathfrak{g}$ be a Lie subalgebra of $\mathfrak{s}_n$,   and
$\mathfrak{g}_i=\pi_i(\mathfrak{g})$ for $i=1,\ldots,n$. If $\mathfrak{g}$
admits algebraically independent and commuting  first integrals $f_1, \ldots,
f_n\in\C(\vx,\vy)$, then
$\mathfrak{g}_{i}$ is Abelian  for $i=1,\ldots, n$.
\end{corollary}
\begin{proposition}
\label{prop:h}
Let $\mathfrak{h}$ be a one dimensional Lie subalgebra of
$\mathrm{sp}(2,\C)\simeq\C_2[x_1,y_1]$. If $f\in\C(x_1,\ldots,
x_n,y_1,\ldots,y_n)\setminus\C$ is a rational first integral of  $\mathfrak{h}$,
then there
exists an element $h\in\C[x_1,y_1]\setminus\C$ such that
$f\in \C(h,x_2,\ldots,x_n,y_2,\ldots,y_n)$.
 \end{proposition}
\begin{proof}
 Let us assume that $\mathfrak{h}$ is nilpotent, i.e., $\mathfrak{h}$ is
generated by $H=y^2$. Then we have
\begin{equation*}
 0=\{H, f\}= -2y_1\pder{f}{x_1}.
\end{equation*}
Thus, $f$ does not depend on $x_1$, so for this case we choose $h=y_1$.

The only other possibility is that  $\mathfrak{h}$ is diagonal, i.e., it
is generated by $H=x_1y_1$. First, let us assume that $f$ is a polynomial in
$(x_1,y_1)$. We can consider $f$ as an element of ring $R[x_1,y_1]$, where
$R=\C(x_2,\ldots,x_n,y_2,\ldots,y_n)$.  We can write $f$ uniquely as a sum of
homogeneous  components.  Here `homogeneity' means the homogeneity with respect
to
$(x_1,y_1)$. It is clear that
if $f$  is a first integral  of $H$, then each homogeneous component of $f$ is
also
a first integral of $H$. Thus, let us assume that $f$ is homogeneous of degree
$s$
and let us represent it in the form 
\begin{equation}
 f =\sum_{i=0}^sf_ix_1^iy_1^{s-i}, \qquad f_i\in R \mtext{for} i=1,\ldots, s.
\end{equation}
Then we obtain
\begin{equation*}
 0=\{H,f\}=
\sum_{i=0}^sf_i(s-i)x_1^iy_1^{s-i}-\sum_{i=0}^sif_ix_1^iy_1^{s-i}=\sum_{i=0}
^sf_i(s-2i)x_1^iy_1^{s-i}.
\end{equation*}
Hence, $f_i=0$ for $2i\neq s$, and if for even $s=2r$, $f_r\neq0$, then
$f=f_r(x_1y_i)^r$. This implies that every homogeneous, and thus arbitrary,
polynomial first integral $f\in R[x_1,y_1]$ of $H$ is an element of
$R[h]$, where $h=x_1y_1$.

Now, assume that $f$ is a rational first integral of $H$. Then we can write
$f=P/Q$ where $P$ and $Q$  are relatively prime polynomials in $R[x_1,y_1]$.
Hence we
have
\begin{equation*}
 0=\{H,P/Q\}=\frac{1}{Q^2}\left( Q\{H,P\}-P\{H,Q\}\right),
\end{equation*}
so $ Q\{H,P\}=P\{H,Q\}$. As $P$ and $Q$ are relatively prime this implies
that
\begin{equation}
\label{eq:rat}
 \{H,P\} = \gamma P \mtext{and} \{H,Q\}=\gamma Q,
\end{equation}
for a certain $\gamma\in R[x_1,y_1]$. 
Comparing the degrees of both sides in the above equalities, we deduce that
$\gamma\in \C$. If $\gamma=0$,
then $P$ and $Q$ are polynomial first integrals of $H$, so in this case we have
that $f\in R(h)=\C(h,x_2,\ldots,x_n,y_2,\ldots,y_n)$.

We show that case $\gamma\neq 0$ is impossible. 
Let us assume that $\gamma\neq 0$. It is easy to see that if $P\in R[x_1,y_1]$
satisfies equation
\begin{equation}
\label{eq:darH}
 \{H,P\} = \gamma P,
\end{equation} 
 then its every  homogeneous component
 also satisfies this equation. Thus let us assume
that $P$ is homogeneous of degree $s$.  If we
write 
\begin{equation}
 P=\sum_{i=0}^sP_ix_1^iy_1^{s-i}, \qquad P_i\in R \mtext{for} i=1,\ldots, s,
\end{equation}
then,   equation~\eqref{eq:darH} leads to the following equality
\begin{equation}
 \sum_{i=0}^sP_i(s-2i-\gamma)x_1^iy_1^{s-i}=0.
\end{equation}
This implies that if coefficient $P_i\neq0$, then $\gamma = s-2i$ and $P= P_i
x_1^i y_1^{s-i}$. Thus,  every homogeneous solution of~\eqref{eq:darH}
is a monomial   of the form $P_i x_1^{i}y^{i+\gamma}$, where $
\gamma$ is a non-zero integer  and $i$ is a non-negative integer such that
$i+\gamma\geq 0$. Thus a non-homogeneous solution of ~\eqref{eq:darH} is a
finite sum
\begin{equation*}
 P = \sum_{i+\gamma>0} p_i x_1^{i}y^{i+\gamma}.
\end{equation*}
But $Q$ satisfies the same equation~\eqref{eq:darH}, so we have also
\begin{equation*}
 Q = \sum_{j+\gamma>0} q_j x_1^{j}y^{j+\gamma}.
\end{equation*}
If $\gamma>0$, then $P$ and $Q$ are not relatively prime because they have a
common factor $y_1^{\gamma}$. On the other hand, if $\gamma<0$, then they are 
not  relatively prime either  because they have a common factor $x_1$. We have a
contradiction  and this finishes the proof.
\end{proof}
\begin{lemma}
\label{lem:com1}
 Let $\mathfrak{g}$ be a Lie subalgebra of $\mathfrak{s}_n$,  
$\mathfrak{g}_i=\pi_i(\mathfrak{g})$ for $i=1,\ldots,n$. Assume that
$f\in\C(\vx,\vy)\setminus\C$ is a 
first integral of  $\mathfrak{g}$. If $\dim_{\C}\mathfrak{g}_i=1$ for
$i=1,\ldots,n$, then there exist $h_i\in\C[x_i,y_i]\setminus\C$ for $i=1,\ldots
,n$ such that $f\in\C(h_1, \ldots, h_n)$.
\end{lemma}
\begin{proof}
It is enough to apply $n$-times Proposition~\ref{prop:h} taking
$\mathfrak{g}_i$ as $\mathfrak{h}$ for $i=1,\ldots,n$.
\end{proof}
\begin{lemma}
\label{lem:comnp1}
 Let $\mathfrak{g}$ be a Lie subalgebra of $\mathfrak{s}_n$,  
$\mathfrak{g}_i=\pi_i(\mathfrak{g})$ for $i=1,\ldots,n$. Assume that
$f_1, \ldots f_{n+1}\in\C(\vx,\vy)$ are algebraically independent
first integrals of  $\mathfrak{g}$ and, moreover, $f_1, \ldots f_{n}$ commute.
Then there exists $1\leq i\leq n$, such that $\dim_{\C}\mathfrak{g}_i=0$.
\end{lemma}
\begin{proof}
 As $\mathfrak{g}$ admits $n$ commuting and independent first integrals,  by
Corollary~\ref{cor:bask},   $\mathfrak{g}_i$ is Abelian, and thus $\dim_{\C}
\mathfrak{g}_i\leq 1$ for $i=1, \ldots,n$.

We prove the statement of the lemma by contradiction. Thus let us assume that 
$\dim_{\C}
\mathfrak{g}_i=1$ for $i=1, \ldots,n$. Then, by Lemma~\ref{lem:com1},
$f_i\in\C(h_1,\ldots, h_n)$ for $i=1,\ldots, n+1$. By assumption $f_1,\ldots,
f_{n+1}$ are algebraically independent. But in $C(h_1, \ldots, h_n)$ any set of
$s>n$ elements is algebraically dependent. A contradiction finishes the proof.
\end{proof}
The above lemma can be generalised in the following way.
\begin{lemma}
\label{lem:comnps}
 Let $\mathfrak{g}$ be a Lie subalgebra of $\mathfrak{s}_n$,  
$\mathfrak{g}_i=\pi_i(\mathfrak{g})$ for $i=1,\ldots,n$. Assume that
$f_1, \ldots f_{n+s}\in\C(\vx,\vy)$, $1\leq s<n$ are algebraically
independent
first integrals of  $\mathfrak{g}$ and, moreover, $f_1, \ldots f_{n}$ commute.
Then there exist $1\leq i_1<\cdots<i_s\leq n$, such that
$\dim_{\C}\mathfrak{g}_{i_j}=0$, for $j=1,\dots, s$.
\end{lemma}
This lemma can be easily proved by induction. We leave a proof to the reader.

\section{Proofs of Theorem~\ref{thm:we1} and~\ref{thm:we2} }
\label{sec:p}
Having the results collected in the two previous sections proofs of
theorems~\ref{thm:we1} and \ref{thm:we2} are very simple.
\begin{proof}[Proof of Theorem~\ref{thm:we1}]
The differential Galois group of variational equations~\eqref{eq:varts} has the
form of product~\eqref{eq:dirsumg}, hence its Lie algebra $\mathfrak{g}$ is a
Lie subalgebra of $\mathfrak{s}_n$. If the considered system admits $l$
functionally independent and commuting first integrals, then by
Proposition~\ref{prop:lieint}, $\mathfrak{g}$ has $l$ algebraically independent
and commuting first integrals. By Lemma~\ref{lem:bask}, there exist $1\leq
i_1<\cdots<i_l\leq n$ such that algebras
$\mathfrak{g}_{i_s}=\pi_{i_s}(\mathfrak{g} )$ are Abelian for $s=1,\ldots, l$. 
As the identity components of the differential Galois groups of
equation~\eqref{eq:solk} and \eqref{eq:hyps} are the same, we  have that
$\mathscr{G}(k, \lambda_{i_s})^\circ$ are Abelian for $s=1,\ldots, l$. The
statement of the theorem follows directly from Proposition~\ref{prop:sol}. 
\end{proof}
\begin{proof}[Proof of Theorem~\ref{thm:we2}]
The normal variational equations for the considered solution are a direct
product of first $n-1$ of equations~\eqref{eq:varts}. Hence, the Lie algebra
$\mathfrak{g}$ of their differential Galois group is a Lie
subalgebra $\mathfrak{s}_{n-1}$. By assumption,  Lie algebra $\mathfrak{g}$
 admits first integrals $f_2,\ldots,f_{n+l}$, such that $(n-1)$ of them
$f_2,\ldots,f_{n}$ commute. By Lemma~\eqref{lem:comnps}, $l$ among Lie algebras
$\mathfrak{g}_i=\pi_i(\mathfrak{s}_{n-1})$, for $i=1,\ldots,n-1$, are zero
dimensional. Without loss of the generality we can assume that
$\dim_{\C}\mathfrak{g}_i=0$ for $i=1,\ldots,l$. Then  $\widehat G(k,
\lambda_{i})^\circ\simeq \mathscr{G}(k, \lambda_{i})^\circ=\{E\}$ for
$i=1,\ldots, l$. Assume that $\abs{k}>2$.  Then, by
point 1. of Corollary~\ref{cor:0}, $(k,\lambda_i)$ belongs to
an item $3$--$9$ in table~\eqref{eq:tab}.  For $\abs{k}\leq 2$, by 
point 2.~of Corollary~\ref{cor:0},  $(k,\lambda_i)$ belongs to an item of
table~\eqref{eq:tab3d}, for $i=1,\ldots,l$.
\end{proof}

\section{Examples}
\label{sec:e}
%-----------------------------------
As the first  example, we consider the following potential 
\begin{equation}
 \label{eq:ck3}
V = \frac{1}{k} \left[ (q_1 - q_2)^k + (q_2 - q_3)^k + (q_3 - q_1)^k \right],
\end{equation} 
with an integer $k$. We exclude the uninteresting cases, $k = 0, 1$,
from the beginning. This system  describes a motion of three
particles with equal masses on a line, with coordinates $q_1, q_2, q_3$,
respectively. For an arbitrary $k$ this system is partially
integrable because it has a first integral 
$$
F_2 = p_1 + p_2 + p_3,
$$
which is the total momentum of the system.
Furthermore, this system is integrable in the Liouville sense for $k \in\{4, 2,
-2\}$,  and even
 super-integrable for $k = \pm 2$. Indeed,  the additional first
integrals for each case are the following:
\begin{itemize}
\item for $k = 4$:
$$
F_3 = p_1(q_2-q_3) + p_2(q_3-q_1) + p_3(q_1-q_2),
$$
\item for $k = 2$:
\begin{equation*}
\begin{split}
F_3 &= p_1(q_2-q_3) + p_2(q_3-q_1) + p_3(q_1-q_2), \\
F_4 &= (p_2-p_3)^2 +3(q_2-q_3)^2,
\end{split}
\end{equation*}
\item for $k = -2$:
\begin{equation*}
\begin{split}
F_3 = &\frac{2}{3}(p_1^3 + p_2^3 + p_3^3) -
\frac{p_1+p_2}{(q_1-q_2)^2}- \frac{p_2+p_3}{(q_2-q_3)^2}-
\frac{p_3+p_1}{(q_3-q_1)^2},\\
F_4 =&
(q_1+q_2+q_3) \left[ p_1^2+p_2^2+p_3^2 - \frac{1}{(q_1-q_2)^2} -
\frac{1}{(q_2-q_3)^2} - \frac{1}{(q_3-q_1)^2} \right] \\
 &\mbox{}- (p_1+p_2+p_3) (p_1 q_1 + p_2 q_2 + p_3 q_3), \\
F_5 =& (p_1 + p_2 + p_3)
\left[ 2(p_1^2 q_1 + p_2^2 q_2 + p_3^2 q_3)  -
\frac{q_1+q_2}{(q_1-q_2)^2}- \frac{q_2+q_3}{(q_2-q_3)^2}-
\frac{q_3+q_1}{(q_3-q_1)^2} \right] \\
&\mbox{}- 3 (q_1+q_2+q_3)
\left[ \frac{2}{3}(p_1^3 + p_2^3 + p_3^3) -
\frac{p_1+p_2}{(q_1-q_2)^2}- \frac{p_2+p_3}{(q_2-q_3)^2}-
\frac{p_3+p_1}{(q_3-q_1)^2} \right].
\end{split}
\end{equation*}
\end{itemize}
Case $k=2$ is just a three particle harmonic oscillator, while case
$k=-2$ is a special case of the Calogero-Moser system. The
Calogero-Moser system  has a Lax pair representation and
this fact allows to prove its super-integrability, see
\cite{Wojciechowski:83::}.
On the other hand, it seems that the case $k=4$ was first realized to
be integrable in \cite{Yoshida:84::a}.
Until now
no further integrals and no further valules of $k$ for which the system is 
integrable or  super-integrable  have been found.

For this system, let us see how Theorems~\ref{thm:we1} and~\ref{thm:we2} work.
For this purpose we
need solutions of the algebraic equation $V'(\vd) =\vd$.  We do not know
how to find all of them  for an arbitrary $k$, nevertheless, it is
sufficient to know some of them. Here we use two solutions. 
 The first one is
$$
 \vd_1 = (c, 0, -c), \qquad  c^{k-2} = \frac{1}{1+ (-1)^k 2^{k-1}}, 
$$
and it exists for all $k\neq 2$. The second one, which exists only when $k>2$ is
an
even integer, is
$$
 \vd_2 = (c, -2c, c), \qquad c^{k-2} = \frac{1}{3^{k-1}}.
$$
The eigenvalues of the Hessian matrix $V''(\vd_1)$ 
are
$$
 (\lambda_{1,1}, \lambda_{1,2}, \lambda_{1,3}) =
\left( \frac{3(k-1)}{1+ (-1)^k 2^{k-1}},\; 0,\; k-1 \right),
$$
and eigenvalues of  $V''(\vd_2)$ are following
$$
 (\lambda_{2,1}, \lambda_{2,2}, \lambda_{2,3}) =
\left( \frac{(k-1)}{3},\; 0,\; k-1 \right).
$$
First, the considered system is partially integrable because of the existence
of 
two commuting first integrals, $F_1 = H$ and $F_2$. Then Theorem 1.2
requires that for each $i=1,2$  at least two pairs of $(k, \lambda_{i,j})$ 
belong to the list~\eqref{eq:tab}. Indeed this is the case, as
$\lambda_{i,2} = 0$ and $\lambda_{i, 3} = k-1$ for $i=1,2$ are always in item 2
of the list~\eqref{eq:tab}.

We show that for values of $k$ different from  given above there is no
integrable cases. 
\begin{lemma}
 \label{lem:ca}
Assume that $k\in\Z\setminus\{-2,0,1,2,4\}$. Then the Hamiltonian system with 
potential~\eqref{eq:ck3} is not integrable in the Liouville sense.
\end{lemma}
\begin{proof} 
We prove the statement of the lemma by a contradiction. Thus let
$k\in\Z\setminus\{-2,0,1,2,4\}$ and the system is integrable in the Liouville
sense. Then, from Theorem~\ref{thm:mo2} or \ref{thm:we1}, it  follows that 
$(k,\lambda_{i,1})$ are  in the list~\eqref{eq:tab}.   We show that for each
$k\in\Z\setminus\{-2,0,1,2,4\}$ either $\lambda_{1,1}$ or $\lambda_{2,1}$ does
not belong to the list.

Assume that $k\geq 3$ is an odd integer. Then $\lambda_{1,1} =
3(k-1)/(1-2^{k-1})<0$ but for positive $k$ the allowed values in the list are
non-negative.

Assume $k \ge 6$ is an even integer. We show that $\lambda_{2,1} =
(k-1)/3$ does not belong to an item of table (1.8). We have only two
possibilities: either $\lambda_{2,1}$ belongs to item 2 or to item 3.
On the other hand, item 2 and item 3 with integer $p$ gives a
strictly increasing sequence of numbers
$$
\left(
 0,\;\; \frac{k-1}{2k},\;\; 1, \;\; k-1,\;\; k+\frac{k-1}{2k},\;\;
k+2, \;\; 3k-2, \;\; 3k+\frac{k-1}{2k}, \;\; \ldots
\right),
$$
while $1 < (k-1)/3 < k-1$ when $k \ge 6$. Thus $\lambda_{2,1}$ does
not belong to an item of table (1.8).

When $k = -1$, $\lambda_{1,1} = -8$ does not belong to the list.

Assume $k \le -3$ is a negative integer. We show that $\lambda_{1,1}
= 3(k-1)/(1+ (-1)^k 2^{k-1})$ does not belong to an item of table
(1.8). We first show that $\lambda_{1,1}$ does not belong to item 2
nor to item 3. Indeed, item 2 and item 3 with integer $p$ gives a
strictly decreasing sequence of numbers
$$
\left(
1, \;\; \frac{k-1}{2k},\;\; 0, \;\; k+2, \;\; k+\frac{k-1}{2k},\;\;
k-1, \;\; 3k+3, \;\; 3k+\frac{k-1}{2k}, \;\; 3k-2, \;\; 6k+ 4, \;\; \ldots
\right).
$$
On the other hand one can easily verify the ineqality
$$
3k-2 > \frac{3(k-1)}{1+ (-1)^k 2^{k-1} } > 6k + 4,
$$
when $k \le -3$. Thus $\lambda_{1,1}$ does not belong to an item 2
and 3 of table (1.8).

 When $k \in \{ -5, -4, -3 \} $, we have to check also that
$\lambda_{1,1}$ does not belong to items 7-9 in table (1.8). This task is
reduced to checking if a certain quadratic
polynomial has an integer root. Let us consider for example case $k=-3$. Then
$\lambda_{1,1}=-64/3$. For $k=-3$ items 2, 3 and 7 are allowed. Assume that
$\lambda_{1,1}$ is given by the first expression in item 7. Then equation
\begin{equation*}
 \dfrac {25} {24}-\dfrac 1 {6}\left( 1 +3p\right)^2= -\frac{64}{3}, 
\end{equation*}
must have an integer solution for $p$. But as it easy to show that it has not
such 
a root.
\end{proof}

Finally, let us apply Theorem~\ref{thm:we2} to potential~\eqref{eq:ck3} and
check how the conditions for the
super-integrability are veryfied. Among
three integrable values of $k$, the case $k=4$ cannot be
super-integrable since the set of eigenvalues is
\begin{equation}
 (\lambda_{1,1}, \lambda_{1,2}, \lambda_{1,3})=(\lambda_{2,1}, \lambda_{2,2},
\lambda_{2,3})=(1, 0, 3)
\end{equation} 
and none of $\lambda_{i,j}$ belongs to items 3-9 of
the list. 
On the other hand,
case $k=-2$ gives $(\lambda_{1,1}, \lambda_{1,2}, \lambda_{1,3})=(-8, 0, -3)$.
The non-trivial eigenvalues are $-8$ and $0$.  These values are compatible with
the 
maximal super-integrability, as  they are given by the first item in
table~\eqref{eq:tab3}. 

For $k=2$, equation $V'(\vd)=\gamma \vd$ has a non-zero solution only for
$\gamma=3$. Eigenvalues $(\hat\lambda_1,\hat\lambda_2,\hat\lambda_3 )$ of
$V''(\vd)$ do not depend on $\vd$, and we have
$(\hat\lambda_1,\hat\lambda_2,\hat\lambda_3 )=(3,0,3)$, so
$(\lambda_1,\lambda_2,\lambda_3 )=(1,0,1)$, see Remark~\ref{rem:k2}. The
nontrivial
eigenvalues are $0$ and $1$. Only one of them, namely 1, belongs to the forth
item in table~\eqref{eq:tab3}, so the system is super-integrable but it cannot
be maximally super-integrable. The fact that the system cannot be maximally
super-integrable also follows from the following. The Hamiltonian of the system
is a quadratic function of canonical variables. Thus we can transform it into
the normal form. It is easy to show that this normal form is following
\begin{equation*}
 K = \frac{1}{2}\sqrt{3}(x_1^2+y_1^2)+ \frac{1}{2}\sqrt{3}(x_2^2+y_2^2) +
\frac{1}{2}y_3^2.
\end{equation*}
Hence the phase curves of the systems lie on two dimensional cylinders. But for
a maximally super-integrable system the maximal dimension of an invariant set
is one.

Potential~\eqref{eq:ck3} has the following higher dimensional generalisation 
\begin{equation}
 \label{eq:ckn}
V= \frac{1}{k}\sum_{i=1}^n (q_i-q_{i+1})^k, \qquad q_{n+1}\equiv q_1.
\end{equation} 
For $k=-2$, as it was shown in \cite{Wojciechowski:83::}, this potential is
maximally super-integrable. The question appears whether it is  integrable for
$k=4$
and $n>3$. We show that for $n=k=4$ this potential is only partially integrable
with no more than one additional first integral.
\begin{lemma}
 \label{lem:n4k4}
Assume that $n=k=4$. Then the Hamiltonian system with potential~\eqref{eq:ckn}
admits no more than two functionally independent meromorphic and commuting first
integrals, namely the Hamiltonian and $F_2=p_1+p_2+p_3+p_4$.
\end{lemma}
\begin{proof}
 For $n=k=4$ algebraic equation $V'(\vd)=\vd$ has a solution
\begin{equation*}
 \vd= \frac{1}{4}(-1,1,1,-1).
\end{equation*}
For this Darboux point eigenvalues of $V''(\vd)$ are following
\begin{equation*}
 (\lambda_1, \lambda_2,\lambda_3,\lambda_4)=\left( \frac{3}{2},\frac{3}{2}, 0,3
\right).
\end{equation*}
It is easy to verify that $\lambda_3$ and $\lambda_4$ are given by the second
item in table~\eqref{eq:tab} but there is no item giving
$\lambda_1=\lambda_2=3/2$. Hence, by Theorem~\eqref{thm:we1}, the system admits
no more than two independent and commuting first integrals. 
\end{proof}
The  classical Bertrand theorem~\cite{05.0470.01}, states that the only radial
potentials
$V(r)=\alpha r^k$, for which all bounded orbits are periodic, are those
with $k=-1$ and $k=2$. The condition that all bounded orbits of a 
Hamiltonian system are periodic means that the system is degenerated, i.e., all 
invariant tori are one dimensional.  Such degeneration appears if the system is
maximally super-integrable. Having this in mind, let us apply
Theorem~\ref{thm:we2}  to a radial potential of the form
\begin{equation}
\label{eq:rk}
 V = \alpha r^k, \qquad \alpha\neq 0, \quad r = \sqrt{q_1^2+q_2^2}, 
\end{equation} 
with an integer $k$. The Hamiltonian  system with this potential is
integrable as it admits  the angular momentum integral, $F_2 = q_1 p_2 - q_2
p_1$. We show the following.
\begin{lemma}
\label{lem:cen}
The Hamiltonian system with potential~\eqref{eq:rk} is super-integrable
if
and only if $k=-1$ and $k=2$. 
\end{lemma}
\begin{proof}
Potential~\eqref{eq:rk} admits  infinitely many  Darboux
points, but, at each of them, the eigenvalues of the Hessian $V''$  are 
$(1, k-1)$, and the non-trivial one is $\lambda = 1$.
Let us apply Theorem~\ref{thm:we2} to check whether this system can be
super-integrable. Assume that $\abs{k}>2$. Then, by Theorem~\ref{thm:we2}, the
non-trivial eigenvalue should be given by an   item $3$--$9$ in
table~\eqref{eq:tab}, but all these items give non-integer values of $\lambda$. 
If $\abs{k}\leq 2$, then $(k,\lambda)$ should belong to an item in
table~\eqref{eq:tab3}. Notice that $(k,\lambda)=(-2,1)$ and $(k,\lambda)=(1,1)$
do not belong to this table while $(k,\lambda)=(2,1)$ and $(k,\lambda)=(-1,1)$
do.
Both cases $k = -1$ %(2D Kepler)
and $k =2$ %(2D isotropic harmonic oscillator)
are indeed super-integrable, because of the existence of the third integral,
$$
F_3 = p_2 (q_1 p_2 - q_2 p_1) + \alpha\frac{q_1}{r},
$$
for $k = -1$, %(a component of Laplace-Runge-Lenz vector),
and
$$
F_3 = p_1 p_2 + 2 \alpha q_1 q_2,
$$
for $k = 2$. 
\end{proof}
Thus, at least as far as the above example is concerned, Theorem~\ref{thm:we1}
and
\ref{thm:we2} together
have the full predicting power. They predict all integrable, partially
integrable and
super-integrable cases without  any exceptions.

\section*{Acknowledgments}
For the second author this research has been partially supported by  the
European Community project
GIFT  (NEST-Adventure Project no. 5006) and by projet de l'Agence National de la
Recherche
"Int\'egrabilit\'e r\'eelle et complexe en m\'ecanique hamiltonienne"
N$^\circ$~JC05$_-$41465.

\renewcommand{\thesection}{\Alph{section}}
\setcounter{section}{0}
\setcounter{equation}{0}

\section{Appendix}
\subsection{Second order differential equations with rational coefficients}
Let us consider a second order
differential equation of the following form
\begin{equation}
\label{eq:gso}
 y''=r y, \qquad r\in\C(z), \qquad '\equiv \frac{\rmd\phantom{z}}{\rmd z} .
\end{equation} 
For this equation its differential Galois group $\mathscr{G}$ is a linear
algebraic
subgroup of $\mathrm{SL}(2,\C)$. The following lemma describes all
possible types of $\mathscr{G}$ and relates these types to the forms of
solutions
of \eqref{eq:gso}, see \cite{Kovacic:86::,Morales:99::c}.
\begin{lemma}
\label{lem:alg}
Let $\mathscr{G}$ be the differential Galois group of equation~\eqref{eq:gso}.
Then one of four cases can occur.
\begin{enumerate}
\item $\mathscr{G}$ is reducible (it is conjugate to a subgroup of triangular
  group); in this case equation \eqref{eq:gso} has an exponential
  solution of the form $y=\exp\int \omega$, where $\omega\in\C(z)$,
\item $\mathscr{G}$ is conjugate with a subgroup of 
\[
\mathscr{DP}= \left\{ \begin{bmatrix} c & 0\\
                                0 & c^{-1}
                      \end{bmatrix}  \; \biggl| \; c\in\C^*\right\} \cup 
                      \left\{ \begin{bmatrix} 0 & c\\
                                c^{-1} & 0
                      \end{bmatrix}  \; \biggl| \; c\in\C^*\right\}, 
\]
  in this case equation
  \eqref{eq:gso} has a solution of the form $y=\exp\int \omega$, where
  $\omega$ is algebraic over $\C(z)$ of degree 2,
\item $\mathscr{G}$ is primitive and finite; in this case all
  solutions of equation \eqref{eq:gso} are algebraic, 
  
\item $\mathscr{G}= \mathrm{SL}(2,\C)$ and equation \eqref{eq:gso}
  has no Liouvillian solution.
\end{enumerate}
\end{lemma}

We  need  a more precise characterisation of case 1 in the above
lemma. It is given by the following lemma, see Lemma~4.2 in
\cite{Singer:93::a}.
\begin{lemma}
\label{lem:algc1}
  Let $\mathscr{G}$ be the differential Galois group of
  equation~\eqref{eq:gso} and assume that $\mathscr{G}$ is reducible.
  Then either
\begin{enumerate}
\item equation~\eqref{eq:gso} has a unique solution $y$ such that
  $y'/y\in\C(z)$, and $\mathscr{G}$ is conjugate to a subgroup of the
  triangular group
\[
 \mathscr{T} = \left\{ \begin{bmatrix} a & b\\
                        0& a^{-1}
               \end{bmatrix} \, |\, a,b\in\C, a\neq 0\right\}. 
\] 
Moreover, $\mathscr{G}$ is a proper subgroup of $\cT$ if and only if there
exists $m\in\N$ such that $y^m\in\C(z)$. In this case $\mathscr{G}$ is
conjugate to
\[
  \mathscr{T}_m = \left\{ \begin{bmatrix} a & b\\
                        0& a^{-1}
               \end{bmatrix} \, |\, a,b\in\C, a^m=1\right\}, 
\]  
where $m$ is the smallest positive integer such that $y^m\in\C(z)$, or
\item equation~\eqref{eq:gso} has two linearly independent solutions
  $y_1$ and $y_2$ such that $y'_i/y_i\in\C(z)$, then $\mathscr{G}$ is
  conjugate to a subgroup of the diagonal group
\[
  \mathscr{D}= \left\{ \begin{bmatrix} a & 0\\
                        0& a^{-1}
               \end{bmatrix} \, |\, a\in\C, a\neq 0\right\}. 
\]
In this case, $y_1y_2\in\C(z)$. Furthermore, $\mathscr{G}$ is conjugate to a
proper subgroup of $ \mathscr{D} $ if and only if $y_1^m\in\C(z)$ for some
$m\in\N$. In this case $\mathscr{G}$ is a cyclic group of order $m$ where $m$
is the smallest positive integer such that $y_1^m\in\C(z)$.
\end{enumerate}
\end{lemma}

\subsection{Riemann $P$ equation}

The Riemann $P$ equation \cite{Whittaker:35::} is the most general
second order differential equation with three regular singularities.
If we place, using homography, these singularities at $z=0$, $z=1$ and
$z=\infty$,
then it has the form
\begin{equation}
\label{eq:riemann}
\begin{split}
\dfrac{\mathrm{d}^2 w}{\mathrm{d}z^2}&+\left(\dfrac{1-\rho_1-\rho_2}{z}+
\dfrac{1-\sigma_1-\sigma_2}{z-1}\right)\dfrac{\mathrm{d} w}{\mathrm{d}z}\\
&+
\left(\dfrac{\rho_1\rho_2}{z^2}+\dfrac{\sigma_1\sigma_2}{(z-1)^2}+
\dfrac{\tau_1\tau_2-\rho_1\rho_2-\sigma_1\sigma_2}{z(z-1)}\right)w=0,
\end{split}
\end{equation}
where $(\rho_1,\rho_2)$, $(\sigma_1,\sigma_2)$ and $(\tau_1,\tau_2)$ are the
exponents at the respective singular points. These exponents satisfy
the Fuchs relation
\[
\sum_{i=1}^2( \rho_i+\sigma_i +\tau_i)=1.
\]
We denote the differences of exponents by
\[
\rho = \rho_1-\rho_2,  \qquad \sigma=\sigma_1-\sigma_2, \qquad \tau=\tau_1
-\tau_2.
\]
The following lemma   gives the necessary and sufficient condition for
\eqref{eq:riemann} to be reducible. It is a classical, well known fact,
see~\cite{Iwasaki:91::}.
\begin{lemma}
 \label{lem:redrie}
Equation~\eqref{eq:riemann} is reducible if and only there exist $i$, $j$, $k\in
\{1,2\}$, such that
\begin{equation}
 \rho_i +\sigma_j +\tau_k \in \Z.
\end{equation}
Equivalently, equation~\eqref{eq:riemann} is reducible if and only if at least
one number among
\begin{equation}
\label{eq:redriediff}
 \rho+\sigma + \tau, \quad - \rho+\sigma + \tau, \quad  \rho-\sigma + \tau,
\quad \rho+\sigma - \tau,
\end{equation}
is an odd integer. 
\end{lemma}
From the above lemma it follows that if equation~\eqref{eq:riemann} is
reducible, then we can always renumber exponents in such a way that
\begin{equation*}
 \rho_1+\sigma_1+\tau_1\in-\N_0, 
\end{equation*}
where $\N_0$ denotes the set of nonnegative integers. But then, from the Fuchs
relation, we also have 
\begin{equation*}
  \rho_2+\sigma_2+\tau_2\in\N.
\end{equation*}
Hence, if~\eqref{eq:riemann} is reducible, we assume from now on that the
exponents
are numbered in this way.

For a more precise characterisation of the monodromy and differential Galois
groups
we need the following two lemmas. The first describes one solution
of~\eqref{eq:riemann} in a case when it is  reducible, see Lemma~4.3.6, p. 90
in~\cite{Iwasaki:91::}.
\begin{lemma}
 \label{lem:iwa1}
Assume that equation~\eqref{eq:riemann} is reducible, and moreover, at lest one
of the exponents' differences
$\rho$, $\sigma$, $\tau$ is not an integer. Then equation~\eqref{eq:riemann}
has a 
solution of the form
\begin{equation*}
 w(z)= z^{\rho_1}(1-z)^{\sigma_1}h(z),
\end{equation*}
where $h(z)$ is a polynomial, and $\deg h(z)\leq n:= - \rho_1-\sigma_1-\tau_1$. 
\end{lemma}
We also need one fact concerning the monodromy group of
equation~\eqref{eq:riemann}. This group is generated by two matrices $M_0$,
$M_1\in\mathrm{GL}(2,\C)$. These matrices correspond to homotopy classes
$[\gamma_0]$ and $[\gamma_1] $ of loops with one common point  encircling once
in the positive sense singularities $z=0$ and $z=1$, respectively. Then we have
the following lemma, see Lemma~4.3.5 on p. 90  in~\cite{Iwasaki:91::}.
\begin{lemma}
 \label{eq:iwa2}
Assume that $M_0$ and $M_1$ are simultaneously diagonalisable. Then at least one
of matrices $M_0$, $M_1$ or $M_0M_1$ is a scalar matrix.
\end{lemma}
If the difference of exponents at a singular point is an integer, then it can
happen that a local solution around this singularity contains a logarithm. Such
a singularity is called logarithmic. In the case of
equation~\eqref{eq:riemann}, it is enough to know the  exponents to decide which
singularity is logarithmic. To formulate the next lemma which gives the
necessary and sufficient conditions for a singularity of \eqref{eq:riemann} 
to be logarithmic we introduce the following notation. For a non-negative
integer $m\in\N_0$ we define 
\begin{equation*}
 \langle m\rangle:=\begin{cases}
 \emptyset & \mtext{if} m=0,\\
\{1, \ldots, m\} &\mtext{otherwise.}
\end{cases}
\end{equation*}
 For
$s\in\{0,1,\infty\}$ let $e_{s,1}$ and  $e_{s,2}$ denote exponents of
equation~\eqref{eq:riemann}, ordered in such a way that $\Re e_{s,1}\geq \Re
e_{s,2}$. With the above notation we have the following.
\begin{lemma}
\label{lem:iwa3}
 Let $r\in\{0,1,\infty\}$. Then $r$ is a
logarithmic singularity of equation~\eqref{eq:riemann} if and only if
$m:=e_{r,1}-e_{r,2}\in \N_0$, and
\begin{equation}
  e_{r,1}+e_{s,i} +e_{t,j}\not\in \langle m\rangle, \mtext{for} i,j\in\{1,2\},
\end{equation}
where $r,s,t$ are pairwise different elements of $\{0,1,\infty\}$.
\end{lemma}
For the proof, see Lemma~4.7  and its proof on pp. 91--93 
in~\cite{Iwasaki:91::}.

For equation \eqref{eq:riemann} the necessary and sufficient
conditions for solvability of the identity component of its
differential Galois group are given by the following theorem due to
Kimura \cite{Kimura:69::}, see also \cite{Morales:99::c}.
\begin{theorem}[Kimura]
\label{thm:kimura}
  The identity component of the differential Galois group of
  equation~\eqref{eq:riemann} is solvable if and only if
\begin{itemize} 
\item[A:] at least one of  four numbers $\rho+\sigma+\tau$, $-\rho+\sigma+\tau$,
$\rho-\sigma+\tau$,  $\rho+\sigma-\tau$,
  is an odd
  integer, or
\item[B:] the numbers $\rho$ or $-\rho$ and $\sigma$ or $-\sigma$ and
  $\tau$ or $-\tau$ belong (in an arbitrary order) to some of the
  following fifteen families
\begin{center} 
\begin{tabular}{|c|c|c|c|c|} 
\hline 
1&$1/2+l$&$1/2+s$&arbitrary complex number&\\\hline 
2&$1/2+l$&$1/3+s$&$1/3+q$&\\\hline 
3&$2/3+l$&$1/3+s$&$1/3+q$&$l+s+q$ even\\\hline 
4&$1/2+l$&$1/3+s$&$1/4+q$&\\\hline 
5&$2/3+l$&$1/4+s$&$1/4+q$&$l+s+q$ even\\\hline 
6&$1/2+l$&$1/3+s$&$1/5+q$&\\\hline 
7&$2/5+l$&$1/3+s$&$1/3+q$&$l+s+q$ even\\\hline 
8&$2/3+l$&$1/5+s$&$1/5+q$&$l+s+q$ even\\\hline 
9&$1/2+l$&$2/5+s$&$1/5+q$&$l+s+q$ even\\\hline 
10&$3/5+l$&$1/3+s$&$1/5+q$&$l+s+q$ even\\\hline 
11&$2/5+l$&$2/5+s$&$2/5+q$&$l+s+q$ even\\\hline 
12&$2/3+l$&$1/3+s$&$1/5+q$&$l+s+q$ even\\\hline 
13&$4/5+l$&$1/5+s$&$1/5+q$&$l+s+q$ even\\\hline 
14&$1/2+l$&$2/5+s$&$1/3+q$&$l+s+q$ even\\\hline 
15&$3/5+l$&$2/5+s$&$1/3+q$&$l+s+q$ even\\\hline 
\end{tabular}\\[1.5ex] 
\end{center} 
where $l,s,q\in\Z$.
\end{itemize} 
\label{kimura} 
\end{theorem} 
If the identity component $G^\circ$ of the differential Galois
group $G$ of equation~\eqref{eq:riemann} is solvable, but the equation  is not
reducible, i.e., if case A in the Kimura theorem does not occur, then the
differential Galois group is either an imprimitive finite group (families
$2$--$15$), or it is a subgroup of $\mathscr{DP}$ group. In the last case $G$
can be finite or whole  $\mathscr{DP}$ group. The following lemma gives a
criterion for distinction of these two cases. 
\begin{lemma}
 \label{lem:dp}
Suppose  equation~\eqref{eq:riemann} is not reducible. Then its differential
Galois group $G$ is a subgroup of $\mathscr{DP}$ group if and only if at two
singular points the differences of exponents are 
half integers. Moreover, $G$ is a finite group if and only if the exponents at
the
remaining singular point are rational. 
\end{lemma}
The above lemma is just case (b) of Theorem~2.9 from~\cite{Churchill:99::a}.

%\bibliographystyle{plainnat}
%\bibliography{mathreva,ajm,yoshida,morales,books,ziglin,moulin,churchill,dgt,%
%oldies,audin,noncommutative,super,mp,rauch,bertrand}

\newcommand{\noopsort}[1]{}\def\cprime{$'$} \def\cprime{$'$} \def\cprime{$'$}
  \def\cprime{$'$} \def\cprime{$'$} \def\cprime{$'$} \def\cprime{$'$}
  \def\cprime{$'$}

\end{document}